\pdfoutput=1 % for arXiv submission using PDFLaTeX engine
\documentclass[%
 reprint,
 amsmath,amssymb,
 aps,
 prl
 longbibliography,
]{revtex4-2}

\usepackage[pdftex]{graphicx} % for arXiv
\usepackage{dcolumn}% Align table columns on decimal point
\usepackage{bm}% bold math
\usepackage[T1]{fontenc}
\newcommand{\bra}{\langle}
\newcommand{\ket}{\rangle}

\newcommand{\parhyphen}[1][1em]{% Symmetric \paragraph hyphen
  {\normalfont\normalsize\bfseries\hspace*{-1em}}\hspace*{#1}---\hspace*{#1}\ignorespaces}
\begin{document}
\footnotetext[1]{Here, we have assumed that the gradient contribution to the internal energy vanishes \cite{Onuki}.}
\footnotetext[2]{See Supplemental Material at [URL will be inserted by publisher] for the derivation of the pressure tensor.}
\footnotetext[3]{For the Euler-Korteweg equations, an Onsager-type condition was obtained in \cite{Debiec_etal_2018}.}
\footnotetext[4]{See Supplemental Material at [URL will be inserted by publisher] for the detailed derivation of (21).}
\footnotetext[5]{See Supplemental Material at [URL will be inserted by publisher] for detailed calculations and explanations.}
\footnotetext[6]{See Supplemental Material at [URL will be inserted by publisher] for the derivation of the equilibrium correlation length.}

% \preprint{APS/123-QED}

\title{Van der Waals Cascade in Supercritical Turbulence near a Critical Point}% Force line breaks with \\

\author{Tomohiro Tanogami}
 % \email{tanogami.tomohiro.84c@st.kyoto-u.ac.jp}
\author{Shin-ichi Sasa}%
\affiliation{%
Department of Physics, Kyoto University, Kyoto 606-8502, Japan
}%

% \collaboration{MUSO Collaboration}%\noaffiliation

% \affiliation{
% \affiliation{
% \affiliation{%
%  Authors' institution and/or address\\
%  This line break forced with \textbackslash\textbackslash
% }%

% \collaboration{CLEO Collaboration}%\noaffiliation

\date{\today}% It is always \today, today,
\begin{abstract}
We investigate a quite strong turbulence in a supercritical fluid near a gas-liquid critical point.
%82
Specifically, we consider a case in which the Kolmogorov scale is much smaller than the equilibrium correlation length $\xi$.
%101
Although equilibrium critical fluctuations are destroyed by turbulence, $\xi$ still provides a crossover length scale between two types of energy cascade.
%128
At scales much larger than $\xi$, the Richardson cascade becomes dominant, whereas at scales much smaller than $\xi$, another type of cascade, which we call the van der Waals cascade, is induced by density fluctuations.
%171
Experimental conditions required to observe the van der Waals cascade are also discussed.
%76
% \begin{description}
% Secondary publications and information retrieval purposes.
% \item[PACS numbers]
% May be entered using the \verb+\pacs{#1}+ command.
% You may use the \texttt{description} environment to structure your abstract;
% use the optional argument of the \verb+\item+ command to give the category of each item. 
% \end{description}
\end{abstract}

\pacs{Valid PACS appear here}% PACS, the Physics and Astronomy

\maketitle
\paragraph*{Introduction.}\parhyphen[0pt]
% Background of fully-developed turbulence
Nonlinearity, which appears ubiquitously in a broad range of phenomena, causes inevitable interference between widely separated time and space scales.
One of the most extreme examples is found in fully developed turbulence.
In turbulence, the kinetic energy is transferred conservatively and continuously from large to small scales in the so-called inertial range \cite{Frisch}. 
The mechanism of this remarkable phenomenon, the Richardson cascade, was intuitively explained by Richardson's depiction of a large vortex splitting into smaller vortices \cite{Richardson_1922}.
As a consequence of the energy transfer, the kinetic energy spectrum exhibits a power-law behavior---the Kolmogorov spectrum \cite{K41_a,K41_c}.
The kinetic energy transported to small scales is dissipated at the Kolmogorov scale, where viscosity begins to predominate, so that the Richardson cascade is inevitably cut off at the Kolmogorov scale.
The important point here is that in standard cases, the Kolmogorov scale is overwhelmingly larger than the microscopic length scales, such as the molecular mean free path \cite{Frisch}.
Therefore, the cascade never reaches microscopic length scales in these cases.

% Introduction of critical phenomena & state problem
Another notable instance in which nonlinearity causes strong interference between widely separated scales is found in critical phenomena.
As an example, in the vicinity of a gas-liquid critical point, the correlation length of equilibrium density fluctuations $\xi$ reaches a macroscopic order of magnitude \cite{Swinney_1973,Cannell_1975}. 
We here consider the strong turbulent regime of a supercritical fluid near a critical point in which $\xi$ is much larger than the Kolmogorov scale.
Even for such strong turbulence, $\xi$ still provides a length scale at which the stress induced by density fluctuations is comparable to the momentum flux. 
In this case, density fluctuations are driven by turbulence, so that the equilibrium critical fluctuations are destroyed.
We then ask how the Richardson cascade is modified by density fluctuations in the turbulence near a critical point. 
Although turbulence in supercritical fluids has been studied over the past few decades, previous studies have focused on cases in which the Kolmogorov scale is larger than $\xi$ \cite{Assenheimer_1993,Ashkenazi_1999high,Ashkenazi_1999spectra,Accary_2009}.

% Model
We answer the above question by studying hydrodynamic equations including density fluctuations.
Specifically, we include a density gradient contribution to the entropy functional to describe the effects of density fluctuations.
Such a formulation that takes into account gradient contributions was originally proposed in the pioneering work of van der Waals \cite{van_der_Waals}, who introduced a gradient term in the Helmholtz free energy density to describe a gas-liquid interface, and the formulation has been widely used in statistical mechanics since the publication of seminal papers by Ginzbrug and Landau for type-I superconductors \cite{Ginzburg_1950} and by Cahn and Hilliard for binary alloys \cite{Cahn_1958}. 
Following the van der Waals theory, Korteweg proposed hydrodynamic equations that contain the \textit{van der Waals stress} (vdW stress), arising from the density gradient \cite{Korteweg,Dunn_Serrin}, and Onuki generalized the theory by including the gradient contribution to both entropy and energy functionals \cite{Onuki_2005,Onuki}.

% Approach
In this study, we analyze the model using a phenomenological approach based on the Onsager ``ideal turbulence'' theory \cite{Onsager_1949,Eyink_Review,Eyink_Sreenivasan}.
The Onsager theory describes the essence of turbulent behavior, such as the Richardson cascade and energy dissipation in the absence of viscosity, the so-called anomalous dissipation \cite{Taylor_1917}.
Although the Onsager theory involves sophisticated mathematical concepts such as weak solutions, it also provides a phenomenological perspective on the relation between cascades and the singularity of the velocity field.
This theory has been recently extended to various turbulent phenomena, such as compressible turbulence \cite{Feireisl_2017,Eyink_Drivas,Drivas_Eyink,Aluie_PRL,Aluie_scale_locality,Aluie_scale_decomposition} and plasma turbulence \cite{Eyink_2018}, and has also been intensively studied from a deep mathematical point of view related to convex integration \cite{De_2013,Isett_2018,Buckmaster_etal_2019,Buckmaster_2019,Szekelyhidi_2019}.

% Result
In this Letter, we show that supercritical turbulence near a critical point exhibits the Richardson cascade and another type of cascade---the \textit{van der Waals cascade}---induced by the van der Waals stress.
First, we derive ``Kolmogorov's $4/5$-law,'' which states that the mean scale-to-scale kinetic energy flux becomes scale-independent in the ``inertial range.''
Second, we show the possibility of the existence of the van der Waals cascade.
Furthermore, we consider the experimental conditions required to observe such \textit{van der Waals turbulence}, which exhibits both the Richardson and van der Waals cascades.

\paragraph*{Setup.}\parhyphen[0pt]
% Setup
Let $\rho$ be the mass density, ${\bf v}$ be the fluid velocity, and $u$ be the internal energy density.
For simplicity, we assume that a fluid is confined in a cube $\Omega=[0,\mathcal{L}]^3$ with periodic boundary conditions.
We further assume that there is no vacuum region; i.e., $\rho({\bf x},t)>0$ for all ${\bf x}\in\Omega$ and $t\in\mathbb{R}$.
Following the van der Waals theory, we include a gradient contribution to the entropy functional to describe enhanced density fluctuations near a critical point \cite{van_der_Waals,Onuki_2002,Onuki_2005,Onuki,Note1}:
\begin{equation}
\mathcal{S}([u],[\rho])=\int_\Omega d^3{\bf x}\left(s(u,\rho)+\dfrac{c(\rho)}{2}|\nabla\rho|^2\right),
\label{entropy}
\end{equation}
where $([u],[\rho]):=(u({\bf x}),\rho({\bf x}))_{{\bf x}\in\Omega}$, $s(u,\rho)$ denotes the entropy density, and $c(\rho)\le0$ is the capillary coefficient.
In the following discussion, we consider the case in which the capillary coefficient is a sufficiently smooth function of $\rho$; e.g., $c(\rho)=\mathrm{const}$ \cite{Onuki}. 
Through thermodynamic relations, the temperature $T(u,\rho)$ and pressure tensor ${\bf P}$ are determined from (\ref{entropy}) \cite{Note2}:
\begin{eqnarray}
{\bf P}&=&p{\bf I}+{\bf \Sigma},
\end{eqnarray}
where $p(u,\rho)$ denotes the pressure defined by $s(u,\rho)$, ${\bf I}$ is the unit tensor, and ${\bf \Sigma}$ is the \textit{vdW stress tensor}, which arises from the gradient contribution and is defined by
\begin{equation}
{\bf \Sigma}:=\left(Tc\rho\Delta\rho+\dfrac{1}{2}Tc'\rho|\nabla\rho|^2+\dfrac{1}{2}Tc|\nabla\rho|^2\right){\bf I}-Tc\nabla\rho\nabla\rho.
\end{equation}
The time evolution of the densities of mass $\rho$, momentum $\rho{\bf v}$, and total energy $\rho|{\bf v}|^2/2+u$ is then governed by the Navier-Stokes-Korteweg equations \cite{Anderson,Gavage,Note3}:
\begin{eqnarray}
\partial_t\rho+\nabla\cdot(\rho{\bf v})=0,
\label{eq:NSK-1}
\end{eqnarray}
\begin{eqnarray}
\partial_t(\rho{\bf v})+\nabla\cdot(\rho{\bf v}{\bf v}+{\bf P}+\bm{\sigma})={\bf f},
\label{eq:NSK-2}
\end{eqnarray}
\begin{eqnarray}
&&\partial_t\left(u+\dfrac{1}{2}\rho|{\bf v}|^2\right)+\nabla\cdot\left\{\left[\left(u+\dfrac{1}{2}\rho|{\bf v}|^2\right){\bf I}+{\bf P}+\bm{\sigma}\right]\cdot{\bf v}\right.\notag\\
&&\left.-\lambda\nabla T\right\}={\bf v}\cdot{\bf f},
\label{eq:NSK-3}
\end{eqnarray}
where ${\bf f}$ denotes an external force acting at large scales $\sim L$, $\lambda$ is the thermal conductivity, and $\bm{\sigma}$ is the viscous stress tensor of the form
\begin{equation}
\sigma_{ij}=-\mu\left(\partial_iv_j+\partial_jv_i-\dfrac{2}{3}\delta_{ij}\nabla\cdot{\bf v}\right)-\zeta\delta_{ij}\nabla\cdot{\bf v}.
\end{equation}
Here $\mu$ and $\zeta$ are the shear and bulk viscosity coefficients, respectively.
We assume that the viscous effect is sufficiently weak for the Kolmogorov scale to be sufficiently smaller than any other length scales.

% Validity of hydrodynamic description
In the following, (\ref{eq:NSK-1})-(\ref{eq:NSK-3}) are applied even to scales smaller than the equilibrium correlation length.
Strictly speaking, dynamics at such scales should be described within the framework of fluctuating hydrodynamics \cite{Onuki_1997}.
In a turbulent regime, however, the equilibrium correlation may be cut off, and the noise terms may be irrelevant for energy transfer.
We therefore assume that (\ref{eq:NSK-1})-(\ref{eq:NSK-3}) are sufficient for our phenomenological argument.

\paragraph*{Characteristic length scales.}\parhyphen[0pt]
% Correlation length
Owing to the effect of the gradient contribution, several characteristic length scales that are not relevant in ordinary fluid turbulence become important.
Let $\rho_0:=\bra\rho\ket$, $c_0:=c(\rho_0)$, and $T_0:=\bra T\ket$ be the typical density, capillary coefficient, and temperature, respectively, where $\bra\cdot\ket$ denotes the volume average $\int_{\Omega}\cdot d^3{\bf x}/\mathcal{L}^3$.
% The sound velocity is characterized by the isothermal compressibility $K_{T_0}:=\rho^{-1}_0\partial\rho(T_0,p)/\partial p$ as $v_0:=(\rho_0K_{T_0})^{-1/2}$, which is zero at a critical point.
In addition, let $v_0:=(\rho_0K_{T_0})^{-1/2}$ be a velocity characterized by the isothermal compressibility $K_{T_0}:=\rho^{-1}_0\partial\rho(T_0,p)/\partial p$, which is zero at a critical point.
One of the most crucial length scales is the correlation length of equilibrium density fluctuations
\begin{eqnarray}
\xi=\dfrac{\sqrt{T_0|c_0|\rho_0}}{v_0},
\end{eqnarray}
which is expressed by the capillary coefficient $c(\rho)$ and parameters in the entropy density $s(u,\rho)$ \cite{Note6}.
The important point here is that even for strong turbulence, $\xi$ still provides a characteristic length scale at which the vdW stress ${\bf \Sigma}$ and momentum flux $\rho{\bf v}{\bf v}$ are comparable.
Let $\ell_c$ be such a length scale.
Using an estimation that $\rho{\bf v}{\bf v}\sim\rho_0v^2_0$ and ${\bf \Sigma}\sim T_0|c_0|\rho^2_0/\ell^2_c$, we obtain
\begin{eqnarray}
% \rho{\bf v}{\bf v}&\sim&{\bf \Sigma}\notag\\
% \rho_0v^2_0&\sim&T_0c_0\rho^2_0/\ell^2_c\notag\\
\ell_c\sim\xi.
\end{eqnarray}
Note that ${\bf \Sigma}$ can be appreciable at small scales because it contains higher-order spatial derivatives.
Therefore, at scales $\gg\ell_c$, the effect of the vdW stress is small compared with the momentum flux, whereas at scales $\ll\ell_c$, the vdW stress becomes relevant.
This observation implies the possibility of the van der Waals cascade, induced by the vdW stress, at scales $\ll\ell_c$.

% Other characteristic length scales
We attempt to seek other characteristic length scales by noting the local kinetic energy balance equation
\begin{eqnarray}
&&\partial_t\left(\dfrac{1}{2}\rho|{\bf v}|^2\right)+\nabla\cdot\left[\left(\dfrac{1}{2}\rho|{\bf v}|^2{\bf I}+{\bf P}+\bm{\sigma}\right)\cdot{\bf v}\right]\notag\\
&&=p\nabla\cdot{\bf v}+{\bf \Sigma}:\nabla{\bf v}+\bm{\sigma}:\nabla{\bf v}+{\bf v}\cdot{\bf f}.
\label{eq:local kinetic energy balance}
\end{eqnarray}
The first term on the right-hand side of (\ref{eq:local kinetic energy balance}), $-p\nabla\cdot{\bf v}$, is the pressure-dilatation, which represents the conversion of kinetic energy into internal energy and vice versa.
Recent numerical simulations \cite{Aluie_2012,Wang-Yang-Shi} suggest that there is a characteristic length scale $\ell_{\mathrm{large}}$ such that the contribution to the global pressure-dilatation $\bra-p\nabla\cdot{\bf v}\ket$ from scales $\gg\ell_{\mathrm{large}}$ is dominant, whereas the contribution from scales $\ll\ell_{\mathrm{large}}$ is negligible.
The second term on the right-hand side of (\ref{eq:local kinetic energy balance}), $-{\bf \Sigma}:\nabla{\bf v}$, which we call the \textit{vdW-stress--strain}, arises because of the gradient contribution.
It also represents the conversion between kinetic and internal energy.
Because both the vdW stress ${\bf \Sigma}$ and strain $\nabla{\bf v}$ change rapidly in space, there may be a characteristic length scale $\ell_{\mathrm{small}}$ such that the contribution to the global vdW-stress--strain $\bra-{\bf \Sigma}:\nabla{\bf v}\ket$ from scales $\gg\ell_{\mathrm{small}}$ is negligible, whereas the contribution from scales $\ll\ell_{\mathrm{small}}$ is dominant.
In the following, we assume the existence of the intermediate asymptotic limit $\ell_{\mathrm{small}}\ll\ell\ll\ell_{\mathrm{large}}$ that satisfies $\ell_\mathrm{small}\ll\ell_c$ and $\ell_c\ll\ell_{\mathrm{large}}$.

\paragraph*{Main result.}\parhyphen[0pt]
% ``Kolmogorov's 4/5-law'
Let $Q^{\mathrm{flux}}_\ell$ be the scale-to-scale kinetic energy flux that represents the energy transfer from scales $>\ell$ to scales $<\ell$:
\begin{eqnarray}
Q^{\mathrm{flux}}_\ell:=\Pi_\ell+\Lambda^{(p)}_\ell+\Lambda^{(\Sigma)}_\ell,
\label{Q}
\end{eqnarray}
where $\Pi_\ell$ is deformation work \cite{A_first_course_in_turbulence}, which corresponds to the energy flux of the Richardson cascade, $\Lambda^{(p)}_\ell$ is baropycnal work \cite{Aluie_PRL,Lees_Aluie}, which arises because of compressibility, and $\Lambda^{(\Sigma)}_\ell$ is \textit{capillary work}, which arises because of the gradient contribution. 
(Precise definitions are given below.)
These three terms represent the energy transfer due to the momentum flux $\rho{\bf v}{\bf v}$, pressure $p$, and vdW stress ${\bf \Sigma}$, respectively.
The first main claim of this Letter concerns the most crucial property of the energy cascade; i.e., in a steady state with a constant mean kinetic energy, $\bra Q^{\mathrm{flux}}_\ell\ket$ becomes scale-independent in the ``inertial range'' $\ell_{\mathrm{small}}\ll\ell\ll\ell_{\mathrm{large}}$:
\begin{eqnarray}
% \bra Q^{\mathrm{flux}}_\ell\ket\approx\epsilon_{\mathrm{eff}}\quad\text{for}\quad\ell_{\mathrm{small}}\ll\ell\ll\ell_{\mathrm{large}},
\bra Q^{\mathrm{flux}}_\ell\ket\approx\epsilon_{\mathrm{eff}},
\label{4/5-law}
\end{eqnarray}
where $\epsilon_{\mathrm{eff}}:=\bra p\nabla\cdot{\bf v}\ket+\bra{\bf v}\cdot{\bf f}\ket$ denotes the effective energy injection rate, which is scale-independent.
We emphasize that, because $Q^{\mathrm{flux}}_\ell$ can be expressed in terms of field increments ${\bf v}({\bf x}+{\bf r})-{\bf v}(\bf x)$ and $\rho({\bf x}+{\bf r})-\rho({\bf x})$, the relation (\ref{4/5-law}) plays the same role as Kolmogorov's $4/5$-law \cite{K41_c}.

% Existence of a double-cascade process
The second main result of this Letter is the prediction of the van der Waals cascade.
In the range of $\ell_c\ll\ell\ll\ell_{\mathrm{large}}$, the Richardson cascade, induced by the momentum flux, becomes dominant, whereas in the range of $\ell_{\mathrm{small}}\ll\ell\ll\ell_c$, the van der Waals cascade, induced by vdW stress, develops:
\begin{eqnarray}
\bra\Lambda^{(\Sigma)}_\ell\ket\ll\bra\Pi_\ell\ket\approx\epsilon_{\mathrm{eff}}\quad\text{for}\quad\ell_c\ll\ell\ll\ell_{\mathrm{large}},\notag\\
\bra\Pi_\ell\ket\ll\bra\Lambda^{(\Sigma)}_\ell\ket\approx\epsilon_{\mathrm{eff}}\quad\text{for}\quad\ell_{\mathrm{small}}\ll\ell\ll\ell_c.
\end{eqnarray}
Correspondingly, the velocity power spectrum exhibits a power-law behavior: $k^{-5/3}$ for $\ell^{-1}_{\mathrm{large}}\ll k\ll\ell^{-1}_c$ and $k^{-3}$ for $\ell^{-1}_c\ll k\ll\ell^{-1}_{\mathrm{small}}$.

\paragraph*{Suggested experiments.}\parhyphen[0pt]
% Suggestion of experiments
We consider the experimental conditions required for observing the van der Waals cascade.
In the study of critical phenomena, $\mathrm{CO}_2$ has been widely used because its critical state occurs under readily realized experimental conditions ($T_c=304.13$ $\mathrm{K}$, $p_c=7.3773$ $\mathrm{MPa}$, $\rho_c=0.4678$ $\mathrm{g}$ $\mathrm{cm}^{-3}$) \cite{Nikolai_2019,Carles_2010}.
In this case, the shear viscosity $\mu$ takes a value around $3.5\times10^{-4}$ $\mathrm{g}$ $\mathrm{cm}^{-1}$ $\mathrm{s}^{-1}$ \cite{Nikolai_2019,Vesovic_1990,Accary_2009}.
We first estimate the Kolmogorov scale $\ell_d$, which can be estimated in terms of $\mu$, $\rho_c$, $L$, and $v_{\mathrm{rms}}:=\sqrt{\bra|{\bf v}|^2\ket}$ as
\begin{eqnarray}
\ell_d\sim L^{1/4}\left(\dfrac{\mu}{\rho_cv_{\mathrm{rms}}}\right)^{3/4}.
\end{eqnarray}
If we achieve a quite strong turbulent regime, in which $\mathrm{Re}\sim10^5$ (e.g., $v_{\mathrm{rms}}\sim100$ $\mathrm{m}/\mathrm{s}$ and $L\sim0.1$ $\mathrm{m}$), the Kolmogorov scale is $\sim800$ \AA.
Therefore, if one can reach the vicinity of the critical point such that the correlation length is at least $\sim10,000$ \AA, it may be possible to verify our predictions by measuring the velocity field using hot-wire anemometry or laser Doppler velocimetry.
To achieve a correlation length of that magnitude, we must control the system with an accuracy of at least $T-T_c\sim10^{-4}$ $\mathrm{K}$ because $\xi\approx\xi_0\epsilon^{-\nu}$, where $\xi_0=1.5$ \AA, $\epsilon:=(T-T_c)/T_c$, and $\nu=0.630$ \cite{Nikolai_2019,Vesovic_1990}.

\paragraph*{Derivation of the main result.}\parhyphen[0pt]
% Coarse-graining
We study the properties of kinetic energy transfer across scales using a coarse-graining approach that can resolve turbulent fields both in scale and in space.
For any field ${\bf a}({\bf x})$, we define a coarse-grained field at length scale $\ell$ as
\begin{equation}
\bar{{\bf a}}_\ell({\bf x}):=\int_{\Omega}d^3{\bf r}G_\ell({\bf r}){\bf a}({\bf x}+{\bf r}),
\label{coarse-graining}
\end{equation}
where $G:\Omega\rightarrow [0,\infty)$ is a smooth symmetric function supported in the open unit ball with $\int_{\Omega}G=1$, and $G_\ell({\bf r}):=\ell^{-3}G({\bf r}/\ell)$ is the rescaling defined for each $\ell>0$.
By coarse-graining (\ref{eq:NSK-1}) and (\ref{eq:NSK-2}), we can write the coarse-grained kinetic energy balance equation
\begin{equation}
\partial_t\left(\dfrac{1}{2}\bar{\rho}_\ell|\tilde{{\bf v}}_\ell|^2\right)+\nabla\cdot{\bf J}_\ell=\bar{p}_\ell\nabla\cdot\bar{{\bf v}}_\ell+\bar{{\bf \Sigma}}_\ell:\nabla\bar{{\bf v}}_\ell-Q^{\mathrm{flux}}_\ell-D_\ell+\epsilon^{\mathrm{in}}_\ell,
\label{eq:coarse-grained kinetic energy balance}
\end{equation}
where we introduce the density-weighted coarse-grained velocity $\tilde{{\bf v}}_\ell:=\overline{(\rho{\bf v})}_\ell/\bar{\rho}_\ell$ to reduce the number of additional cumulant terms and to obtain a simple physical interpretation.
Here, $\epsilon^{\mathrm{in}}_\ell:=\tilde{{\bf v}}_\ell\cdot\bar{{\bf f}}_\ell$ denotes the energy injection rate due to external stirring at scale $\ell$, $D_\ell:=-\nabla\tilde{{\bf v}}_\ell:\bar{{\bm \sigma}}_\ell$ denotes the viscous dissipation acting at scale $\ell$, and ${\bf J}_\ell$ represents the spatial transport of large-scale kinetic energy, which does not contribute to the energy transfer across scales.
The first two terms on the right-hand side of (\ref{eq:coarse-grained kinetic energy balance}), $-\bar{p}_\ell\nabla\cdot\bar{{\bf v}}_\ell$ and $-\bar{{\bf \Sigma}}_\ell:\nabla\bar{{\bf v}}_\ell$, are the large-scale pressure-dilatation and vdW-stress--strain, respectively. 
Note that these two terms contain no modes at small scales $<\ell$.
Therefore, they contribute only to the conversion of the large-scale kinetic energy into internal energy and vice versa. 
The third term on the right-hand side of (\ref{eq:coarse-grained kinetic energy balance}) denotes the scale-to-scale kinetic energy flux (\ref{Q}).
The definitions and physical meanings of each term comprising $Q^{\mathrm{flux}}_\ell$ are given as follows.
Deformation work is defined by
\begin{equation}
\Pi_\ell:=-\bar{\rho}_\ell\nabla\tilde{{\bf v}}_\ell:\tilde{\tau}_\ell({\bf v},{\bf v}),
\end{equation}
where $\tilde{\tau}_\ell({\bf v},{\bf v}):=\widetilde{({\bf v}{\bf v})}_\ell-\tilde{{\bf v}}_\ell\tilde{{\bf v}}_\ell$, and it represents the work done by the large-scale ($>\ell$) strain $\nabla\tilde{{\bf v}}_\ell$ against the small-scale ($<\ell$) stress $\bar{\rho}_\ell\tilde{\tau}_\ell({\bf v},{\bf v})$.
Baropycnal work is defined by
\begin{equation}
\Lambda^{(p)}_\ell:=\dfrac{1}{\bar{\rho}_\ell}\nabla\bar{p}_\ell\cdot\bar{\tau}_\ell(\rho,{\bf v}),
\end{equation}
where $\bar{\tau}_\ell(\rho,{\bf v}):=\overline{(\rho{\bf v})}_\ell-\bar{\rho}_\ell\bar{{\bf v}}_\ell$, and it represents the work done by the large-scale pressure gradient force $-\nabla\bar{p}_\ell/\bar{\rho}_\ell$ against the small-scale mass flux $\bar{\tau}_\ell(\rho,{\bf v})$.
Capillary work, which has a form similar to that of baropycnal work,
\begin{equation}
\Lambda^{(\Sigma)}_\ell:=\dfrac{1}{\bar{\rho}_\ell}\nabla\cdot\bar{{\bf \Sigma}}_\ell\cdot\bar{\tau}_\ell(\rho,{\bf v}),
\end{equation}
represents the work done by the large-scale force $\nabla\cdot\bar{{\bf \Sigma}}_\ell/\bar{\rho}_\ell$ against the small-scale mass flux $\bar{\tau}_\ell(\rho,{\bf v})$.
Note that in (\ref{eq:coarse-grained kinetic energy balance}), only these three terms are capable of the direct transfer of kinetic energy across scales because each of the three terms has a form ``large-scale ($>\ell$) quantity $\times$ small-scale ($<\ell$) quantity,'' whereas the other terms on the right-hand side of (\ref{eq:coarse-grained kinetic energy balance}) do not.

% Derivation of the main result
In the steady state, the spatial averaging of (\ref{eq:coarse-grained kinetic energy balance}) gives
\begin{eqnarray}
\bra Q^{\mathrm{flux}}_\ell\ket=\bra\bar{p}_\ell\nabla\cdot\bar{{\bf v}}_\ell\ket+\bra\bar{{\bf \Sigma}}_\ell:\nabla\bar{{\bf v}}_\ell\ket-\bra D_\ell\ket+\bra\epsilon^{\mathrm{in}}_\ell\ket.
\label{eq:Q}
\end{eqnarray}
For the first term on the right-hand side, because $\bra p\nabla\cdot{\bf v}\ket$ receives most of its contribution from scales $\gg\ell_{\mathrm{large}}$, it can be approximated as $\bra\bar{p}_\ell\nabla\cdot\bar{{\bf v}}_\ell\ket\approx\bra p\nabla\cdot{\bf v}\ket$ for $\ell\ll\ell_{\mathrm{large}}$.
Similarly, because the contribution to $\bra{\bf \Sigma}:\nabla{\bf v}\ket$ from scales $\gg\ell_{\mathrm{small}}$ is negligible, the second term becomes $\bra\bar{{\bf \Sigma}}_\ell:\nabla\bar{{\bf v}}_\ell\ket\approx0$ for $\ell\gg\ell_{\mathrm{small}}$.
In addition, because the Kolmogorov scale is sufficiently smaller than other length scales and ${\bf f}$ acts at the large scale $L$, the viscous dissipation $\bra\bar{D}_\ell\ket$ and energy injection $\bra\epsilon^{\mathrm{in}}_\ell\ket$ can be approximated as $\bra\bar{D}_\ell\ket\approx0$ and $\bra\epsilon^{\mathrm{in}}_\ell\ket\approx\bra{\bf v}\cdot{\bf f}\ket$ for $\ell_{\mathrm{small}}\ll\ell\ll\ell_{\mathrm{large}}$, respectively \cite{Aluie_scale_decomposition}.
Thus, in the intermediate scale range $\ell_{\mathrm{small}}\ll\ell\ll\ell_{\mathrm{large}}$, (\ref{eq:Q}) becomes \cite{Note4}
\begin{eqnarray}
\bra Q^{\mathrm{flux}}_\ell\ket&\approx&\bra p\nabla\cdot{\bf v}\ket+\bra{\bf v}\cdot{\bf f}\ket\notag\\
&=&\epsilon_{\mathrm{eff}}.
\label{eq:Q_inertial_range}
\end{eqnarray} 
Note that although the mean total scale-to-scale kinetic energy flux $\bra Q^{\mathrm{flux}}_\ell\ket$ is scale-independent in the inertial range, the three terms $\bra\Pi_\ell\ket$, $\bra\Lambda^{(p)}_\ell\ket$, and $\bra\Lambda^{(\Sigma)}_\ell\ket$ are not necessarily scale-independent individually.
In fact, because ${\bf \Sigma}$ is appreciable at scales $\ll\ell_c$, $\bra\Pi_\ell\ket$ and $\bra\Lambda^{(p)}_\ell\ket$ are dominant at scales $\gg\ell_c$, whereas $\bra\Lambda^{(\Sigma)}_\ell\ket$ develops at scales $\ll\ell_c$.
Therefore, (\ref{eq:Q_inertial_range}) can be further rewritten as
\begin{equation}
\bra Q^{\mathrm{flux}}_\ell\ket\approx
\begin{cases}
\bra\Pi_\ell\ket+\bra\Lambda^{(p)}_\ell\ket\approx\epsilon_{\mathrm{eff}}\quad\text{for}\quad\ell_c\ll\ell\ll\ell_{\mathrm{large}},\\
\bra\Lambda^{(\Sigma)}_\ell\ket\approx\epsilon_{\mathrm{eff}}\quad\text{for}\quad\ell_{\mathrm{small}}\ll\ell\ll\ell_c.
\end{cases}
\label{eq:Q_double-cascade}
\end{equation}
We can also derive the result (\ref{eq:Q_double-cascade}) more rigorously by evaluating the scale dependence of the energy fluxes using functional analysis \cite{Note5}.

% Energy spectrum---large scale
We now consider the singularity of the velocity field that is necessary to satisfy (\ref{eq:Q_double-cascade}).
To this end, we investigate $\delta a(\ell):=|{\bf a}({\bf x}+{\bm \ell})-{\bf a}({\bf x})|$ for a field ${\bf a}({\bf x})$ using the assumption of homogeneity and isotropy.
The following pointwise estimation is based on a more sophisticated analysis using Besov spaces \cite{Note5}.
In the range of $\ell_c\ll\ell\ll\ell_{\mathrm{large}}$, the baropycnal work and deformation work are the main sources of the energy cascade.
These two energy fluxes can be expressed in terms of increments:
\begin{equation}
% \Lambda^{(p)}_\ell=\dfrac{1}{\bar{\rho}_\ell}\nabla\bar{p}_\ell\cdot\bar{\tau}_\ell(\rho,{\bf v})\sim\dfrac{\bar{p}_L\delta\rho(\ell)\delta v(\ell)}{\rho_0L}\rightarrow0\quad\text{as}\quad\ell/L\rightarrow0,
\Lambda^{(p)}_\ell=\dfrac{1}{\bar{\rho}_\ell}\nabla\bar{p}_\ell\cdot\bar{\tau}_\ell(\rho,{\bf v})\sim\dfrac{\delta p(\ell)\delta\rho(\ell)\delta v(\ell)}{\rho_0\ell},
\label{eq:Lambda^p_estimation}
\end{equation}
\begin{equation}
\Pi_\ell=-\bar{\rho}_\ell\nabla\tilde{{\bf v}}_\ell:\tilde{\tau}_\ell({\bf v},{\bf v})\sim-\dfrac{\rho_0(\delta v(\ell))^3}{\ell},
\label{eq:Pi_estimation}
\end{equation}
where we have used $\bar{\rho}_\ell\sim\rho_0$ assuming homogeneity.
We have also used an estimation that $\nabla\bar{f}_\ell\sim\delta f(\ell)/\ell$ and $\bar{\tau}_\ell(f,g)\sim\delta f(\ell)\delta g(\ell)$, which can be made more rigorous using the $L^p$-norm \cite{Note5}.
From this expression, it follows that $\bra\Lambda^{(p)}_\ell\ket\rightarrow0$ as $\ell\rightarrow0$ because the density increment is bounded from above as $\delta\rho(\ell)=O(\ell)$, which holds for all $\ell>0$ because the entropy functional contains the density gradient term $\propto|\nabla\rho|^2<\infty$.
We therefore conclude that $\bra\Pi_\ell\ket\approx\epsilon_{\mathrm{eff}}$ for $\ell_c\ll\ell\ll\ell_{\mathrm{large}}$.
Then, from the expression (\ref{eq:Pi_estimation}), we obtain $\delta v(\ell)\sim\rho^{-1/3}_0\epsilon^{1/3}_{\mathrm{eff}}\ell^{1/3}$ in this scale range.
This result implies that the velocity power spectrum $E^v(k)$ follows the Kolmogorov spectrum: $E^v(k)\sim\rho^{-2/3}_0\epsilon^{2/3}_{\mathrm{eff}}k^{-5/3}$ for $\ell^{-1}_{\mathrm{large}}\ll k\ll\ell^{-1}_c$.

% Energy spectrum---small scale
In the range of $\ell_{\mathrm{small}}\ll\ell\ll\ell_c$, the energy transfer is dominated by capillary work.
The capillary work can be expressed in terms of increments:
\begin{eqnarray}
\Lambda^{(\Sigma)}_\ell=\dfrac{1}{\bar{\rho}_\ell}\nabla\cdot\bar{{\bf \Sigma}}_\ell\cdot\bar{\tau}_\ell(\rho,{\bf v})\sim\dfrac{\bar{\Sigma}_\ell\delta\rho(\ell)\delta v(\ell)}{\rho_0\ell}.
\end{eqnarray}
Here, note that we cannot naively estimate as in $\nabla\cdot\bar{{\bf \Sigma}}_\ell\sim\delta\Sigma(\ell)/\ell$ because ${\bf \Sigma}$ already contains higher-order derivatives of $\rho$.
In this scale range, we must consider the density increment $\delta\rho(\ell)$ because density fluctuations are appreciable.
The scale dependence of $\delta\rho(\ell)$ can be complicated because of the strong turbulent effect \cite{Hnativc_2018}, although it may be bounded from above as $\delta\rho(\ell)=O(\ell)$.
Here, as a first step to estimate the spectral exponent, we consider the consequence of imposing only the loose condition that $\delta\rho(\ell)=O(\ell)$.
Then, by integrating by parts, we can estimate that $\bar{\Sigma}_\ell\delta\rho(\ell)\sim Z$, where $Z$ is a scale-independent quantity \cite{Note5}.
Hence, we conclude that $\delta v(\ell)\sim Z^{-1}\rho_0\epsilon_{\mathrm{eff}}\ell$ and $E^v(k)\sim Z^{-2}\rho^2_0\epsilon^2_{\mathrm{eff}}k^{-3}$ in this scale range.

\paragraph*{Concluding remarks.}\parhyphen[0pt]
% Summary
In summary, we have shown that supercritical turbulence near a critical point can exhibit the van der Waals cascade.
The interesting point here is that the results are similar to those reported for pure quantum turbulence \cite{Tanogami_2020}.
This implies that pure quantum turbulence and van der Waals turbulence belong to the same ``universality class.''
Therefore, our results may also provide an interesting perspective from which to understand quantum turbulence, which will help illuminate the role of quantized vortices and Kelvin waves.
Finally, we remark that there is a possibility that the spectrum $k^{-3}$ becomes shallower because of the depletion of nonlinearity \cite{Eyink_2018,Boldyrev_2005,Mason_etal_2006,Tanogami_2020} or the regularity of the temperature and density gradient fields \cite{Note5}.

% Importance of experiments
The problem addressed in this Letter could lead to an understanding not only of turbulence but also of the relation between the macroscopic and microscopic descriptions of nature.
We therefore hope that experiments will be conducted to verify our predictions.

% \begin{acknowledgements}
The present study was supported by JSPS KAKENHI Grant No. 20J20079, a Grant-in-Aid for JSPS Fellows, and JSPS KAKENHI (Nos. 17H01148, 19H05795, and 20K20425).
% \end{acknowledgements}

% \bibliographystyle{apsrev4-2}
% \bibliographystyle{jabbrv}
\bibliography{main_text}

\end{document}

\onecolumngrid
% \clearpage
%%%%%%%%%% Prefix a "S" to all equations, figures, tables and reset the counter %%%%%%%%%%
\setcounter{equation}{0}
\setcounter{figure}{0}
\setcounter{table}{0}
\makeatletter
\renewcommand{\theequation}{S\arabic{equation}}
\renewcommand{\thefigure}{S\arabic{figure}}
% \renewcommand{\bibnumfmt}[1]{[S#1]}
% \renewcommand{\citenumfont}[1]{S#1}
%%%%%%%%%% Prefix a "S" to all equations, figures, tables and reset the counter %%%%%%%%%%
\appendix
% \section*{Supplemental Material}
\title{Supplemental Material: Van der Waals Cascade in Supercritical Turbulence near a Critical Point}% Force line breaks with \\

\author{Tomohiro Tanogami}
 % \email{tanogami.tomohiro.84c@st.kyoto-u.ac.jp}
\author{Shin-ichi Sasa}%
\affiliation{%
Department of Physics, Kyoto University, Kyoto 606-8502, Japan
}%

% \collaboration{MUSO Collaboration}%\noaffiliation

% \affiliation{
% \affiliation{
% \affiliation{%
%  Authors' institution and/or address\\
%  This line break forced with \textbackslash\textbackslash
% }%

% \collaboration{CLEO Collaboration}%\noaffiliation

\date{\today}% It is always \today, today,
% \begin{abstract}
% \begin{description}
% Secondary publications and information retrieval purposes.
% \item[PACS numbers]
% May be entered using the \verb+\pacs{#1}+ command.
% You may use the \texttt{description} environment to structure your abstract;
% use the optional argument of the \verb+\item+ command to give the category of each item. 
% \end{description}
% \end{abstract}

\pacs{Valid PACS appear here}% PACS, the Physics and Astronomy

\maketitle
\onecolumngrid

\subsection{Derivation of the pressure tensor}
Here, we derive the pressure tensor ((2) and (3) in the main text).
The equilibrium value of $([u],[\rho])=(u({\bf x}),\rho({\bf x}))_{{\bf x}\in \Omega}$ in the isolated system enclosed by adiabatic walls, denoted as $(u_*({\bf x}),\rho_*({\bf x}))$, is determined as the maximizer of the entropy functional,
\begin{equation}
\mathcal{S}([u],[\rho])=\int_\Omega d^3{\bf x}\left(s(u,\rho)+\dfrac{c(\rho)}{2}|\nabla\rho|^2\right).
\label{entropy}
\end{equation}
It follows the conservation law,
\begin{equation}
\int_\Omega d^3{\bf x}\rho({\bf x})/m=N,
\end{equation}
\begin{equation}
\int_\Omega d^3{\bf x}u({\bf x})=E,
\end{equation}
where $E$ and $N$ are constants.
The variational equation is
\begin{eqnarray}
\dfrac{1}{T(u_*,\rho_*)}=\lambda_1,
\end{eqnarray}
\begin{eqnarray}
-\dfrac{\mu(u_*,\rho_*)}{T(u_*,\rho_*)}+\dfrac{c'(\rho_*)}{2}|\nabla\rho_*|^2-\nabla\cdot(c(\rho_*)\nabla\rho_*)=\lambda_2,
\label{eq:mu}
\end{eqnarray}
where $\lambda_1$ and $\lambda_2$ are Lagrange multipliers that are physically connected to the equilibrium values of temperature and chemical potential as $\lambda_1=1/T^{\mathrm{eq}}$ and $\lambda_2=-\mu^{\mathrm{eq}}/T^{\mathrm{eq}}$, respectively.

We define
\begin{equation}
\tilde{\mu}:=\mu(u,\rho)-T(u,\rho)\left(\dfrac{c'(\rho)}{2}|\nabla\rho|^2-\nabla\cdot\left(c(\rho)\nabla\rho\right)\right),
\label{eq:generalized_mu}
\end{equation}
such that the equilibrium condition is given by $\nabla\tilde{\mu}={\bf 0}$.
We then determine $\tilde{p}$, such that $\nabla\cdot\tilde{p}={\bf 0}$ in equilibrium and $\tilde{p}=p(u,\rho)$ when the gradient terms are ignored.
To this end, we use a relation,
\begin{equation}
\nabla(p/T)=-u\nabla(1/T)+\rho\nabla(\mu/T),
\label{Gibbs-Duhem}
\end{equation}
which is derived from
\begin{eqnarray}
p=\mu\rho-u+Ts,
\end{eqnarray}
\begin{eqnarray}
\nabla s=\dfrac{1}{T}\nabla u-\dfrac{\mu}{T}\nabla\rho.
\end{eqnarray}
We first rewrite the second term on the right-hand side of (\ref{Gibbs-Duhem}) in terms of the generalized chemical potential, $\tilde{\mu}$, as
\begin{eqnarray}
\rho\nabla(\mu/T)&=&\rho\nabla(\tilde{\mu}/T)-\rho\nabla((\tilde{\mu}-\mu)/T)\notag\\
&=&\rho\nabla(\tilde{\mu}/T)-\nabla(\rho(\tilde{\mu}-\mu)/T)+\nabla\cdot\left(-\dfrac{c}{2}|\nabla\rho|^2{\bf I}+c\nabla\rho\nabla\rho\right)\notag\\
&=&\rho\nabla(\tilde{\mu}/T)-\nabla\cdot\left(\rho c\Delta\rho{\bf I}+\dfrac{1}{2}\rho c'|\nabla\rho|^2{\bf I}+\dfrac{1}{2}c|\nabla\rho|^2{\bf I}-c\nabla\rho\nabla\rho\right).
\label{eq:rho nabla mu T}
\end{eqnarray}
Here, we used the relation,
\begin{eqnarray}
(\tilde{\mu}-\mu)/T=-\dfrac{1}{2}c'|\nabla\rho|^2+\nabla\cdot(c\nabla\rho),
\end{eqnarray}
which follows from the definition of $\tilde{\mu}$ (\ref{eq:generalized_mu}).
By substituting this result into (\ref{Gibbs-Duhem}), we obtain
\begin{equation}
\nabla(\tilde{p}/T)=-u\nabla(1/T)+\rho\nabla(\tilde{\mu}/T)
\end{equation}
with
\begin{eqnarray}
\tilde{p}=\left(p+T\rho c\Delta\rho+\dfrac{1}{2}T\rho c'|\nabla\rho|^2+\dfrac{1}{2}Tc|\nabla\rho|^2\right){\bf I}-Tc\nabla\rho\nabla\rho.
\label{generalized p}
\end{eqnarray}
The equilibrium condition, $\nabla T={\bf 0}$ and $\nabla\tilde{\mu}={\bf 0}$, leads to $\nabla\cdot\tilde{p}={\bf 0}$.
In addition, it is evident that $\tilde{p}=p{\bf I}$ when the density gradient is ignored.
In the main text, we used the notation ${\bf P}=\tilde{p}$ to emphasize that $\tilde{p}$ is a second-order tensor and defined the van der Waals stress ${\bf \Sigma}$ as
\begin{eqnarray}
{\bf \Sigma}&:=&{\bf P}-p{\bf I}\notag\\
&=&\left(Tc\rho\Delta\rho+\dfrac{1}{2}Tc'\rho|\nabla\rho|^2+\dfrac{1}{2}Tc|\nabla\rho|^2\right){\bf I}-Tc\nabla\rho\nabla\rho.
\end{eqnarray}

\subsection{Correlation length of equilibrium density fluctuations}
In this section, we derive the correlation length of equilibrium density fluctuations and thus confirm that the correlation length is determined by the capillary coefficient and parameters in the entropy density.
To this end, we introduce the Helmholtz free energy functional,
\begin{eqnarray}
\mathcal{F}(T,[\rho]):=\int_\Omega d^3{\bf x}\left(f(T,\rho)-\dfrac{1}{2}Tc(\rho)|\nabla\rho|^2\right),
\label{Helmholtz}
\end{eqnarray}
where $f:=u-Ts$.
Assuming small, slowly varying deviations in density, we consider the expansion of $f$ in terms of the local deviation, $\delta\rho({\bf x}):=\rho({\bf x})-\rho_0$, as follows:
\begin{eqnarray}
f(T,\rho)=f(T,\rho_0)+\mu(T,\rho_0)\delta\rho+\dfrac{1}{2}\dfrac{1}{\rho^2_0K_T}(\delta\rho)^2+\cdots,
\label{free energy density expansion}
\end{eqnarray}
where $K_T$ is the isothermal compressibility, given by
\begin{eqnarray}
K_T:=\dfrac{1}{\rho}\left.\dfrac{\partial \rho(T,p)}{\partial p}\right|_{\rho=\rho_0}.
\end{eqnarray}
Substituting (\ref{free energy density expansion}) into (\ref{Helmholtz}), we obtain
\begin{eqnarray}
\mathcal{F}(T,[\rho])&\simeq&\int_\Omega d^3{\bf x}\left(f(T,\rho_0)+\dfrac{1}{2}\dfrac{1}{\rho^2_0K_T}(\delta\rho)^2-\dfrac{1}{2}Tc_0|\nabla\rho|^2\right)\notag\\
&=&\int_\Omega d^3{\bf x}f(T,\rho_0)+\delta\mathcal{F}(T,[\rho]),
\end{eqnarray}
where 
\begin{eqnarray}
\delta\mathcal{F}(T,[\rho]):=\int_\Omega d^3{\bf x}\left(\dfrac{1}{2}\dfrac{1}{\rho^2_0K_T}(\delta\rho)^2-\dfrac{1}{2}Tc_0|\nabla\rho|^2\right).
\label{free energy deviation}
\end{eqnarray}
Here, the first power of $\delta\rho$ has been dropped considering the conservation of particles, and $c(\rho)$ is replaced by $c_0:=c(\rho_0)$ because the difference $c(\rho)-c_0$ is a higher-order contribution.

Introducing the Fourier transform of the density deviation,
\begin{eqnarray}
\delta\hat{\rho}({\bf k})=\dfrac{1}{V}\int_{\Omega}d^3{\bf x}e^{-i{\bf k}\cdot{\bf x}}\delta\rho({\bf x}),
\end{eqnarray}
where $V:=\mathcal{L}^3$ and ${\bf k}\in(2\pi/\mathcal{L})\mathbb{Z}$, (\ref{free energy deviation}) becomes
\begin{eqnarray}
\delta\mathcal{F}(T,[\rho])=\dfrac{1}{2}V\sum_{\bf k}\left(\dfrac{1}{\rho^2_0K_T}-Tc_0k^2\right)|\delta\hat{\rho}({\bf k})|^2,
\label{free energy deviation k}
\end{eqnarray}
and $k:=|{\bf k}|$.
% If we regard (\ref{free energy deviation k}) as the effective Hamiltonian, the probability distribution of $\{\Delta\hat{\rho}({\bf k})\}_{\bf k}$ is
% \begin{eqnarray}
% \mathcal{P}(\{\Delta\hat{\rho}({\bf k})\}_{\bf k})=\dfrac{1}{Z}\exp\left(-\dfrac{V}{2k_{\mathrm{B}}T}\sum_{\bf k}\left(\dfrac{1}{\rho^2_0K_T}-Tc_0k^2\right)|\Delta\hat{\rho}({\bf k})|^2\right),
% \end{eqnarray}
% where $Z$ is the partition function obtained by integration over all possible configurations
% \begin{eqnarray}
% Z&=&\prod_{\bf k}\int d\Delta\hat{\rho}({\bf k})\exp\left(-\dfrac{V}{2k_{\mathrm{B}}T}\sum_{\bf k}\left(\dfrac{1}{\rho^2_0K_T}-Tc_0k^2\right)|\Delta\hat{\rho}({\bf k})|^2\right)\notag\\
% &=&\prod_{\bf k}Z_{\bf k},
% \end{eqnarray}
% where
% \begin{eqnarray}
% Z_{\bf k}:=\int d\Delta\hat{\rho}({\bf k})\exp\left(-\dfrac{V}{2k_{\mathrm{B}}T}\sum_{\bf k}\left(\dfrac{1}{\rho^2_0K_T}-Tc_0k^2\right)|\Delta\hat{\rho}({\bf k})|^2\right).
% \end{eqnarray}
According to fluctuation theory in equilibrium statistical mechanics, $\delta F(T,[\rho])$ plays a role of an effective Hamiltonian describing density fluctuations of the system with temperature $T$.
That is, the density correlation function takes the Ornstein-Zernike form \cite{Fisher_1964}, as follows:
\begin{eqnarray}
\bra|\delta\hat{\rho}({\bf k})|^2\ket&=&\dfrac{{\displaystyle \int} \left(\prod_{\bf q}d\delta\hat{\rho}({\bf q})\right)|\delta\hat{\rho}({\bf k})|^2\exp\left(-\dfrac{V}{2k_{\mathrm{B}}T}\sum_{\bf q}\left(\dfrac{1}{\rho^2_0K_T}-Tc_0q^2\right)|\delta\hat{\rho}({\bf q})|^2\right)}{{\displaystyle \int} \left(\prod_{\bf q}d\delta\hat{\rho}({\bf q})\right)\exp\left(-\dfrac{V}{2k_{\mathrm{B}}T}\sum_{\bf q}\left(\dfrac{1}{\rho^2_0K_T}-Tc_0q^2\right)|\delta\hat{\rho}({\bf q})|^2\right)}\notag\\
&=&\dfrac{k_{\mathrm{B}}T}{V}\dfrac{1}{(\rho^2_0K_T)^{-1}-Tc_0k^2}\notag\\
&=&\dfrac{k_{\mathrm{B}}}{VT|c_0|}\dfrac{1}{\xi^{-2}+k^2}\quad\text{for}\quad{\bf k}\neq{\bf 0}.
\end{eqnarray}
Here, $\xi$ is the correlation length of density fluctuations
\begin{eqnarray}
\xi&:=&\sqrt{T|c_0|\rho^2_0K_T}\notag\\
&=&\dfrac{\sqrt{T|c_0|\rho_0}}{v_0},
\label{xi}
\end{eqnarray}
where we introduce a velocity characterized by the isothermal compressibility, as follows:
\begin{eqnarray}
v_0:=\dfrac{1}{\sqrt{\rho_0K_T}}.
\end{eqnarray}

As an example, we consider a van der Waals fluid for which the equation of state is given as follows:
\begin{eqnarray}
% p(T,n)=\dfrac{k_{\mathrm{B}}Tn}{1-bn}-an^2,
p(T,\rho)=\dfrac{k_{\mathrm{B}}T}{m}\dfrac{\rho}{1-b\rho}-a\rho^2,
\end{eqnarray}
where $m$ denotes the mass of a particle; the heat capacity per unit volume is given by
\begin{eqnarray}
% c_V(T,n)=\eta k_{\mathrm{B}}n,
c_V(T,\rho)=\eta k_{\mathrm{B}}\rho,
\end{eqnarray}
where $a$, $b$, and $\eta$ are constants.
In this case, the entropy density is given by
\begin{eqnarray}
% s(u,n)=k_{\mathrm{B}}n\log\dfrac{1-bn}{n}+\eta k_{\mathrm{B}}n\log\dfrac{u+an^2}{n}+cn,
s(u,\rho)=\dfrac{k_{\mathrm{B}}}{m}\rho\log\dfrac{1-b\rho}{\rho/m}+\eta k_{\mathrm{B}}\rho\log\dfrac{u+a\rho^2}{\rho/m}+c\rho,
\label{van der Waals entropy density}
\end{eqnarray}
where $c$ is a constant.
The critical density, temperature, and pressure are expressed as
\begin{eqnarray}
\rho_c=\dfrac{1}{3b},\quad T_c=\dfrac{8am}{27bk_{\mathrm{B}}},\quad p_c=\dfrac{a}{27b^2},
\label{critical values}
\end{eqnarray}
respectively.
If $\rho_0=\rho_c$, the isothermal compressibility can be expressed as
\begin{eqnarray}
% K_T=\dfrac{1}{\rho}\left(\dfrac{k_{\mathrm{B}}T}{m}\dfrac{1}{1-b\rho}+\dfrac{k_{\mathrm{B}}T}{m}\dfrac{b\rho}{(1-b\rho)^2}-2a\rho\right)^{-1}.
K_T=\dfrac{1}{6p_c}\dfrac{T_c}{T-T_c}.
\label{isothermal compressibility_van de Waals}
\end{eqnarray}
From (\ref{xi}), (\ref{van der Waals entropy density}), (\ref{critical values}), and (\ref{isothermal compressibility_van de Waals}), it is straightforward to confirm that the correlation length $\xi$ is determined by the capillary coefficient $c(\rho)$ and the parameters in the entropy density $s(u,\rho)$.

\subsection{Preliminaries}
In this section, in preparation for the detailed derivation and explanation of the main result, we introduce the Besov regularity and investigate the scale dependence of energy fluxes and vdW-stress--strain.

\subsubsection{Besov regularity}
To investigate the scale dependence of the scale-to-scale kinetic energy fluxes, we assume that the following scaling laws hold for the absolute structure functions in the inviscid limit for $p\in[1,\infty]$:
% \begin{eqnarray}
% \bra|\delta{\bf v}({\bf r};{\bf x})|^p\ket\sim A^p_pv^p_0\left(\dfrac{|{\bf r}|}{L}\right)^{\zeta_p}\notag\\
% \bra|\delta\rho({\bf r};{\bf x})|^p\ket\sim B^p_p\rho^p_0\left(\dfrac{|{\bf r}|}{L}\right)^{\zeta^\rho_p}\notag\\
% \bra|\delta p({\bf r};{\bf x})|^p\ket\sim C^p_pp^p_0\left(\dfrac{|{\bf r}|}{L}\right)^{\zeta^p_p},
% \end{eqnarray}
\begin{eqnarray}
\|\delta{\bf v}({\bf r};\cdot)\|_p\sim A_pv_0\left(\dfrac{|{\bf r}|}{L}\right)^{\sigma_p},\label{assumption_v}\\
\|\delta\rho({\bf r};\cdot)\|_p=O\left(\left(\dfrac{|{\bf r}|}{L}\right)^{\sigma^\rho_p}\right),\label{assumption_rho}\\
\|\delta p({\bf r};\cdot)\|_p=O\left(\left(\dfrac{|{\bf r}|}{L}\right)^{\sigma^p_p}\right),\label{assumption_p}
\end{eqnarray}
where $A_p$ is a dimensionless constant, $\|\cdot\|_p:=\bra|\cdot|^p\ket^{1/p}$ is the $L^p$-norm, and $\delta{\bf a}({\bf r};{\bf x}):={\bf a}({\bf x}+{\bf r})-{\bf a}({\bf x})$ for any field ${\bf a}({\bf x})$.
The symbol $\sim$ denotes ``asymptotically equivalence,'' i.e., $f(x)\sim g(x)$ for $x\rightarrow0$ if and only if $\lim_{x\rightarrow0}f(x)/g(x)=1$.
% $\|\cdot\|_p$ is the $L^p$-norm, i.e., 
% \begin{eqnarray}
% \|{\bf a}\|_p&:=&\left(\dfrac{1}{\mathcal{L}^3}\int_{\Omega}|{\bf a}({\bf x})|^pd^3{\bf x}\right)^{1/p}\notag\\
% &=&\bra|{\bf a}|^p\ket^{1/p}
% \end{eqnarray} 
% for $p\in[0,\infty)$, and
% \begin{equation}
% \|{\bf a}\|_\infty:=\inf I_{{\bf a}}=\mathrm{ess}\sup|{\bf a}|
% \end{equation}
% for $p=\infty$, where $I_{{\bf a}}$ denotes the semi-infinite interval
% \begin{equation}
% I_{{\bf a}}:=\{c\in[0,\infty)|\text{the set $\{|{\bf a}|>c\}$ is of measure zero}\}.
% \end{equation}
% For a matrix $A=(a_{ij})$, we define its $L^p$-norm $\|A\|_p=\bra|A|^p\ket^{1/p}$ via the Frobenius norm $|A|:=\sqrt{\sum_i\sum_j|a_{ij}|^2}$.
Hereafter, we consider only the case that $0<\sigma_p\le1$ and $0<\sigma^p_p\le1$, because this requirement is well-established empirically in ordinary fluid turbulence \cite{Frisch,Gotoh,Hill,Xu}.
%\textbf{More reference (see Aluie)}.
We also assume that $\sigma^\rho_p=1$, which is a reasonable requirement, because the entropy functional contains the density gradient term, $\propto|\nabla\rho|^2<\infty$.
We note that $\|\delta{\bf v}({\bf r};\cdot)\|_p$ is essentially the traditional absolute structure function $S_p=\bra|\delta{\bf v}({\bf r};\cdot)|^p\ket=\|\delta{\bf v}({\bf r};\cdot)\|^p_p$, and that the scaling relations (\ref{assumption_v})-(\ref{assumption_p}) correspond to the Besov regularity \cite{Eyink_1995,Perrier_1996}.

\subsubsection{Scale dependence of energy fluxes and vdW-stress--strain\label{Scale dependence of energy fluxes and vdW-stress--strain}}
In this section, we study the scale dependence of deformation work, baropycnal work, capillary work, and the vdW-stress--strain.

\paragraph*{Deformation work.}
We now examine the scale $\ell$ dependence of the deformation work, $\Pi_\ell=-\bar{\rho}_\ell\nabla\tilde{{\bf v}}_\ell:\tilde{\tau}_\ell({\bf v},{\bf v})$.
Using the Cauchy-Schwarz and H\"older inequalities, we obtain
\begin{eqnarray}
\|\Pi_\ell\|_{p/3}&=&\|\bar{\rho}_\ell\nabla\tilde{{\bf v}}_\ell:\tilde{\tau}_\ell({\bf v},{\bf v})\|_{p/3}\notag\\
&\le&\|\rho\|_\infty\|\nabla\tilde{{\bf v}}_\ell\|_p\|\tilde{\tau}_\ell({\bf v},{\bf v})\|_{p/2},
\label{Pi_factors}
\end{eqnarray}
where $\|A\|_p=\bra|A|^p\ket^{1/p}$ for a matrix $A=(a_{ij})$ is defined using the Frobenius norm, that is, $|\nabla\tilde{{\bf v}}_\ell|:=\sqrt{\sum^3_{i=1}\sum^3_{j=1}|\partial_i\widetilde{(v_j)}_\ell({\bf x})|^2}$ and $|\tilde{\tau}_\ell({\bf v},{\bf v})|:=\sqrt{\sum^3_{i=1}\sum^3_{j=1}|\tilde{\tau}_\ell(v_i,v_j)|^2}$.

For the second factor on the right-hand side of (\ref{Pi_factors}), $\|\nabla\tilde{{\bf v}}_\ell\|_p$, if we use the relation
\begin{equation}
\tilde{{\bf v}}_\ell=\bar{{\bf v}}_\ell+\dfrac{\bar{\tau}_\ell(\rho,{\bf v})}{\bar{\rho}_\ell}
\label{eq:tilde_u decompose}
\end{equation}
and the Minkowski inequality, we obtain
\begin{eqnarray}
\|\nabla\tilde{{\bf v}}_\ell\|_p&=&\left\|\nabla\left(\bar{{\bf v}}_\ell+\dfrac{\bar{\tau}_\ell(\rho,{\bf v})}{\bar{\rho}_\ell}\right)\right\|_p \notag\\
&\le&\left\|\nabla\bar{{\bf v}}_\ell\right\|_p+\left\|\dfrac{1}{\bar{\rho}_\ell}\nabla\bar{\tau}_\ell(\rho,{\bf v})\right\|_p+\left\|\dfrac{\bar{\tau}_\ell(\rho,{\bf v})}{\bar{\rho}^2_\ell}\nabla\bar{\rho}_\ell\right\|_p.
\label{eq:nabla tilde u}
\end{eqnarray}
Considering the first term on the right-hand side of (\ref{eq:nabla tilde u}), it should be noted that, for any locally integrable function ${\bf a}({\bf x})$,
\begin{equation}
\nabla\bar{{\bf a}}_\ell({\bf x})=-\dfrac{1}{\ell}\int_{\Omega}d^3{\bf r}(\nabla G)_\ell({\bf r})\delta{\bf a}({\bf r};{\bf x})
\end{equation}
because $\int d^3{\bf r}\nabla G({\bf r})={\bf 0}$.
Subsequently, the triangle inequality gives
\begin{eqnarray}
\|\nabla\bar{{\bf a}}_\ell\|_p&=&\left\|\dfrac{1}{\ell}\int_{\Omega}d^3{\bf r}(\nabla G)_\ell({\bf r})\delta{\bf a}({\bf r};\cdot)\right\|_p \notag\\
&\le&\dfrac{1}{\ell}\int_{\Omega}d^3{\bf r}|(\nabla G)_\ell({\bf r})|\|\delta{\bf a}({\bf r};\cdot)\|_p \notag\\
&\le&\dfrac{\mathrm{(const)}}{\ell}\sup_{|{\bf r}|<\ell}\|\delta{\bf a}({\bf r};\cdot)\|_p.
\label{eq:estimate_nabla_a}
\end{eqnarray}
Hence,
\begin{equation}
\|\nabla\bar{{\bf v}}_\ell\|_p=O\left(\dfrac{\|\delta{\bf v}(\ell)\|_p}{\ell}\right),
\label{eq:estimate_nabla_v_tilde_1}
\end{equation}
where $\|\delta{\bf a}(\ell)\|_p:=\sup_{|{\bf r}|<\ell}\|\delta{\bf a}({\bf r};\cdot)\|_p$.
For the second and last terms of (\ref{eq:nabla tilde u}), using Propositions 3 and 4 in \cite{Drivas_Eyink}, we can obtain
\begin{eqnarray}
\left\|\dfrac{1}{\bar{\rho}_\ell}\nabla\bar{\tau}_\ell(\rho,{\bf v})\right\|_p&\le&\dfrac{\mathrm{(const)}}{\ell}\|1/\bar{\rho}_\ell\|_\infty\|\delta\rho(\ell)\|_\infty\|\delta{\bf v}(\ell)\|_p \notag\\
&\le&\dfrac{\mathrm{(const)}}{\ell}\|\delta\rho(\ell)\|_\infty\|\delta{\bf v}(\ell)\|_p,
\label{eq:estimate_nabla_v_tilde_2}
\end{eqnarray}
and
\begin{eqnarray}
\left\|\dfrac{\bar{\tau}_\ell(\rho,{\bf v})}{\bar{\rho}^2_\ell}\nabla\bar{\rho}_\ell\right\|_p&\le&\|1/\bar{\rho}_\ell\|^2_\infty\|\nabla\bar{\rho}_\ell\|_\infty\|\bar{\tau}_\ell(\rho,{\bf v})\|_p \notag\\
&\le&\dfrac{\mathrm{(const)}}{\ell}\|\delta\rho(\ell)\|^2_\infty\|\delta{\bf v}(\ell)\|_p.
% &\le&\mathrm{(const)}M^2\|\rho\|^2_\infty\dfrac{\|\delta{\bf v}(\ell)\|_p}{\ell}.
\label{eq:estimate_nabla_v_tilde_3}
\end{eqnarray}
Thus, combining the results (\ref{eq:estimate_nabla_v_tilde_1}), (\ref{eq:estimate_nabla_v_tilde_2}), and (\ref{eq:estimate_nabla_v_tilde_3}), we obtain
\begin{eqnarray}
\|\nabla\tilde{{\bf v}}_\ell\|_p&=&\dfrac{\|\delta{\bf v}(\ell)\|_p}{\ell}\Bigl[O(1)+O(\|\delta\rho(\ell)\|_\infty)+O(\|\delta\rho(\ell)\|^2_\infty)\Bigr] \notag\\
&=&O\left(\dfrac{\|\delta {\bf v}(\ell)\|_p}{\ell}\right).
\label{eq:estimate nabla_tilde_u}
\end{eqnarray}

For the last factor on the right-hand side of (\ref{Pi_factors}), $\|\tilde{\tau}_\ell({\bf v},{\bf v})\|_{p/2}$, if we use the relation
\begin{equation}
\tilde{\tau}_\ell({\bf v},{\bf v})=\bar{\tau}_\ell({\bf v},{\bf v})+\dfrac{1}{\bar{\rho}_\ell}\bar{\tau}_\ell(\rho,{\bf v},{\bf v})-\dfrac{1}{\bar{\rho}^2_\ell}\bar{\tau}_\ell(\rho,{\bf v})\bar{\tau}_\ell(\rho,{\bf v})
\label{eq:tilde_u_u decompose}
\end{equation}
and the Minkowski inequality, we obtain
\begin{eqnarray}
\|\tilde{\tau}_\ell({\bf v},{\bf v})\|_{p/2}&\le&\|\bar{\tau}_\ell({\bf v},{\bf v})\|_{p/2}+\left\|\dfrac{1}{\bar{\rho}_\ell}\bar{\tau}_\ell(\rho,{\bf v},{\bf v})\right\|_{p/2}+\left\|\dfrac{1}{\bar{\rho}^2_\ell}\bar{\tau}_\ell(\rho,{\bf v})\bar{\tau}_\ell(\rho,{\bf v})\right\|_{p/2}.
\end{eqnarray}
Subsequently, using Proposition 3 in \cite{Drivas_Eyink}, we obtain
\begin{eqnarray}
\|\tilde{\tau}_\ell({\bf v},{\bf v})\|_{p/2}&=&\|\delta{\bf v}(\ell)\|^2_p\Bigl[O(1)+O(\|\delta\rho(\ell)\|_\infty)+O(\|\delta\rho(\ell)\|^2_\infty)\Bigr] \notag\\
&=&O\left(\|\delta{\bf v}(\ell)\|^2_p\right), \quad p\ge2.
\label{eq:estimate tilde_tau_u_u}
\end{eqnarray}

Thus, from (\ref{Pi_factors}), (\ref{eq:estimate nabla_tilde_u}), (\ref{eq:estimate tilde_tau_u_u}), and condition (\ref{assumption_v}), we finally obtain
\begin{eqnarray}
\|\Pi_\ell\|_{p/3}&=&\|\bar{\rho}_\ell\nabla\tilde{{\bf v}}_\ell:\tilde{\tau}_\ell({\bf v},{\bf v})\|_{p/3}\notag\\
&=&O\left(\dfrac{\|\delta{\bf v}(\ell)\|^3_p}{\ell}\right) \notag\\
&=&O\left(\left(\dfrac{\ell}{L}\right)^{3\sigma_p-1}\right),\quad p\ge3,
\label{eq:Estimation of the second term}
\end{eqnarray}
as a rigorous upper bound.
Note that the upper bound of (\ref{eq:Estimation of the second term}) becomes independent of $\ell$ in the case of $\sigma_p=1/3$.

\paragraph*{Baropycnal work.}
Next, we study the scale $\ell$ dependence of the baropycnal work, $\Lambda^{(p)}_\ell=(1/\bar{\rho}_\ell)\nabla\bar{p}_\ell\cdot\bar{\tau}_\ell(\rho,{\bf v})$.
Using the Cauchy-Schwarz and H\"older inequalities, we obtain 
\begin{eqnarray}
\|\Lambda^{(p)}_\ell\|_{p/3}&=&\|(1/\bar{\rho}_\ell)\nabla\bar{p}_\ell\cdot\bar{\tau}_\ell(\rho,{\bf v})\|_{p/3}\notag\\
&\le&\|1/\bar{\rho}_\ell\|_\infty\|\nabla\bar{p}_\ell\|_p\|\bar{\tau}_\ell(\rho,{\bf v})\|_{p/2}.
\end{eqnarray}
For $\|\nabla\bar{p}_\ell\|_p$, from the inequality (\ref{eq:estimate_nabla_a}), we obtain
\begin{eqnarray}
\|\nabla\bar{p}_\ell\|_p=O\left(\dfrac{\|\delta p(\ell)\|_p}{\ell}\right).
\end{eqnarray}
For $\|\bar{\tau}_\ell(\rho,{\bf v})\|_{p/2}$, using Proposition 3 in \cite{Drivas_Eyink}, we obtain
\begin{equation}
\|\bar{\tau}_\ell(\rho,{\bf v})\|_{p/2}=O(\|\delta\rho(\ell)\|_p\|\delta{\bf v}(\ell)\|_p).
\label{eq:estimate tau_rho_u}
\end{equation}
From the requirements (\ref{assumption_v}), (\ref{assumption_rho}), and (\ref{assumption_p}), we obtain
\begin{eqnarray}
\|\Lambda^{(p)}_\ell\|_{p/3}&=&\|(1/\bar{\rho}_\ell)\nabla\bar{p}_\ell\cdot\bar{\tau}_\ell(\rho,{\bf v})\|_{p/3} \notag\\
&=&O\left(\dfrac{1}{\ell}\|\delta p(\ell)\|_p\|\delta\rho(\ell)\|_p\|\delta{\bf v}(\ell)\|_p\right) \notag\\
&=&O\left(\left(\dfrac{\ell}{L}\right)^{\sigma_p+\sigma^p_p}\right), \quad p\ge3.
\label{eq:Estimation of baropycnal work}
\end{eqnarray}
This result implies that the mean baropycnal work, $\bra\Lambda^{(p)}_\ell\ket\le\bra|\Lambda^{(p)}_\ell|\ket=\|\Lambda^{(p)}_\ell\|_1$, vanishes as $O((\ell/L)^{\sigma_3+\sigma^p_3})$ for $\ell/L\rightarrow0$.
Therefore, the baropycnal work does not contribute to the transfer of kinetic energy across scales.

\paragraph*{Capillary work\label{Capillary work}.}
We now investigate the scale $\ell$ dependence of the capillary work, $\Lambda^{(\Sigma)}_\ell=(1/\bar{\rho}_\ell)\nabla\cdot\bar{{\bf \Sigma}}_\ell\cdot\bar{\tau}_\ell(\rho,{\bf v})$.
From the Cauchy-Schwarz and H\"older inequalities, we obtain
\begin{eqnarray}
\|\Lambda^{(\Sigma)}_\ell\|_{p/3}&=&\left\|(1/\bar{\rho}_\ell)\nabla\cdot\bar{{\bf \Sigma}}_\ell\cdot\bar{\tau}_\ell(\rho,{\bf v})\right\|_{p/3}\notag\\
&\le&\|1/\bar{\rho}_\ell\|_\infty\|\nabla\overline{(Tc\rho\Delta\rho)}_\ell\|_p\|\bar{\tau}_\ell(\rho,{\bf v})\|_{p/2}+\|1/\bar{\rho}_\ell\|_\infty\|\nabla\cdot\bar{{\bf \Xi}}_\ell\|_p\|\bar{\tau}_\ell(\rho,{\bf v})\|_{p/2},
\label{Lambda^Sigma_Cauchy-Schwarz}
\end{eqnarray}
where ${\bf \Xi}$ is defined by
\begin{equation}
{\bf \Xi}:=\left(\dfrac{1}{2}Tc'\rho|\nabla\rho|^2+\dfrac{1}{2}Tc|\nabla\rho|^2\right){\bf I}-Tc\nabla\rho\nabla\rho.
\end{equation}

$\|\nabla\overline{(Tc\rho\Delta\rho)}_\ell\|_p$ can be evaluated as follows:
Because
\begin{eqnarray}
\nabla\overline{(Tc\rho\Delta\rho)}_\ell&=&-\dfrac{1}{\ell}\int_\Omega d^3{\bf r}(\nabla G)_\ell({\bf r})(Tc\rho\Delta\rho)({\bf x}+{\bf r})\notag\\
&=&\dfrac{1}{\ell^2}\int_\Omega d^3{\bf r}(\nabla\nabla G)_\ell({\bf r})\cdot(Tc\rho\nabla\rho)({\bf x}+{\bf r})+\dfrac{1}{\ell}\int_\Omega d^3{\bf r}(\nabla G)_\ell({\bf r})(\nabla(Tc\rho)\cdot\nabla\rho)({\bf x}+{\bf r}),\notag\\
\end{eqnarray}
we obtain
\begin{eqnarray}
\|\nabla\overline{(Tc\rho\Delta\rho)}_\ell\|_p&\le&\dfrac{1}{\ell^2}\int_\Omega d^3{\bf r}|(\nabla\nabla G)_\ell({\bf r})|\|Tc\rho\nabla\rho\|_p+\dfrac{1}{\ell}\int_\Omega d^3{\bf r}|(\nabla G)_\ell({\bf r})|\|\nabla(Tc\rho)\cdot\nabla\rho\|_p\notag\\
&=&O\left(\ell^{-2}\right).
\label{estimation_Sigma_1}
\end{eqnarray}

Similarly, $\|\nabla\cdot\bar{{\bf \Xi}}_\ell\|$ can be evaluated as
\begin{eqnarray}
\|\nabla\cdot\bar{{\bf \Xi}}_\ell\|_p&\le&\dfrac{1}{\ell}\int_\Omega d^3{\bf r}|(\nabla G)_\ell({\bf r})|\|{\bf \Xi}\|_p\notag\\
&=&O\left(\ell^{-1}\right).
\label{estimation_Sigma_2}
\end{eqnarray}

Therefore, from (\ref{Lambda^Sigma_Cauchy-Schwarz}), (\ref{eq:estimate tau_rho_u}), (\ref{estimation_Sigma_1}), (\ref{estimation_Sigma_2}), and the conditions (\ref{assumption_v}) and (\ref{assumption_rho}), we obtain
\begin{eqnarray}
\|\Lambda^{(\Sigma)}_\ell\|_{p/3}&=&\|(1/\bar{\rho}_\ell)\nabla\cdot\bar{{\bf \Sigma}}_\ell\cdot\bar{\tau}_\ell(\rho,{\bf v})\|_{p/3}\notag\\
&=&O\left(\left(\dfrac{\ell}{L}\right)^{\sigma_p-1}\right),\quad p\ge3.
\label{estimation_Capillary_work}
\end{eqnarray}
Note that the scale-independent upper bound is obtained in the case of $\sigma_p=1$.

\paragraph*{VdW-stress--strain.}
Finally, we investigate the scale $\ell$ dependence of the large-scale vdW-stress--strain $-\bar{{\bf \Sigma}}_\ell:\nabla\bar{{\bf v}}_\ell$.
From the Cauchy-Schwarz and H\"older inequalities, we obtain
\begin{eqnarray}
\|\bar{{\bf \Sigma}}_\ell:\nabla\bar{\bf v}_\ell\|_p\le\|\bar{{\bf \Sigma}}_\ell\|_\infty\|\nabla\bar{\bf v}_\ell\|_p.
\end{eqnarray}
From a similar argument as (\ref{estimation_Sigma_1}) and (\ref{estimation_Sigma_2}), it follows that
\begin{eqnarray}
\|\bar{{\bf \Sigma}}_\ell\|_\infty&=&\left\|\overline{(Tc\rho\Delta\rho)}_\ell{\bf I}+\bar{{\bf {\bf \Xi}}}_\ell\right\|_\infty\notag\\
&\le&\left\|-\dfrac{1}{\ell}\int_\Omega d^3{\bf r}(\nabla G)_\ell({\bf r})\cdot(Tc\rho\nabla\rho)(\cdot+{\bf r}){\bf I}\right\|_\infty+\left\|\int_\Omega d^3{\bf r}G_\ell({\bf r}){\bf \Xi}(\cdot+{\bf r})\right\|_\infty\notag\\
&\le&\dfrac{\sqrt{3}}{\ell}\int_\Omega d^3{\bf r}|(\nabla G)_\ell({\bf r})|\|Tc\rho\nabla\rho\|_\infty+\int_\Omega d^3{\bf r}|G_\ell({\bf r})|\|{\bf \Xi}\|_\infty\notag\\
&=&O\left(\ell^{-1}\right).
\label{eq:estimation_Sigma}
\end{eqnarray}
% \begin{eqnarray}
% |\bar{{\bf \Sigma}}_\ell|&=&\left|\overline{(Tc\rho\Delta\rho)}_\ell{\bf I}+\bar{{\bf {\bf \Xi}}}_\ell\right|\notag\\
% &\le&\left|-\dfrac{1}{\ell}\int_\Omega d^3{\bf r}(\nabla G)_\ell({\bf r})\cdot(Tc\rho\nabla\rho)_\ell({\bf x}+{\bf r}){\bf I}\right|+\left|\int_\Omega d^3{\bf r}G_\ell({\bf r}){\bf \Xi}({\bf x}+{\bf r})\right|\notag\\
% &\le&\dfrac{\sqrt{3}}{\ell}\int_\Omega d^3{\bf r}|(\nabla G)_\ell({\bf r})||(Tc\rho\nabla\rho)_\ell({\bf x}+{\bf r})|+\int_\Omega d^3{\bf r}|G_\ell({\bf r})||{\bf \Xi}({\bf x}+{\bf r})|\notag\\
% &=&O\left(\ell^{-1}\right).
% \label{eq:estimation_Sigma}
% \end{eqnarray}
Therefore, using the inequality (\ref{eq:estimate_nabla_v_tilde_1}), we obtain
\begin{eqnarray}
\|\bar{{\bf \Sigma}}_\ell:\nabla\bar{{\bf v}}_\ell\|_p=O\left(\left(\dfrac{\ell}{L}\right)^{\sigma_p-2}\right).
\label{estimation_Koretweg-stress-strain}
\end{eqnarray}

% From the assumption (\ref{assumption_rho_naive}) and the estimation (\ref{eq:estimation_Sigma}), 
% \begin{equation}
% \bar{{\bf \Sigma}}_\ell\delta\rho(\ell;{\bf x})=O(1).
% \end{equation}

\subsection{Detailed derivation of ``Kolmogorov's $4/5$-law''}
% In this section, we explain the detailed derivation of the ``Kolmogorov's $4/5$-law.''
In the steady state, spatial averaging of the coarse-grained kinetic energy balance gives
\begin{eqnarray}
\bra Q^{\mathrm{flux}}_\ell\ket=\bra\bar{p}_\ell\nabla\cdot\bar{{\bf v}}_\ell\ket+\bra\bar{{\bf \Sigma}}_\ell:\nabla\bar{{\bf v}}_\ell\ket-\bra D_\ell\ket+\bra\epsilon^{\mathrm{in}}_\ell\ket.
\label{eq:Q}
\end{eqnarray}
Next, we determine the scale range such that the right-hand side of (\ref{eq:Q}) becomes scale-independent.

First, we can prove that the viscous dissipation term, $\bra D_\ell\ket$, can be ignored at scales that are much larger than the Kolmogorov scale, which is sufficiently smaller than other length scales \cite{Aluie_scale_decomposition}.
In addition, because the external force, ${\bf f}$, acts at the large scale $L$, it follows that \cite{Aluie_scale_decomposition}
\begin{eqnarray}
\bra\epsilon^{\mathrm{in}}_\ell\ket&:=&\bra\tilde{{\bf v}}_\ell\cdot\bar{{\bf f}}_\ell\ket\notag\\
&\approx&\bra{\bf v}\cdot{\bf f}\ket\quad\text{for}\quad\ell\ll L.
% &\approx&\bra{\bf v}\cdot{\bf f}\ket-\bra({\bf v}-\tilde{{\bf v}}_\ell)\cdot{\bf f}\ket
\end{eqnarray}

Next, we show that $\bra\bar{p}_\ell\nabla\cdot\bar{{\bf v}}_\ell\ket\approx\bra p\nabla\cdot{\bf v}\ket$ for $\ell\ll\ell_{\mathrm{large}}$.
In the main text, $\ell_{\mathrm{large}}$ is introduced as the characteristic length scale such that the contribution to the global pressure-dilatation $\bra-p\nabla\cdot{\bf v}\ket$ from scales much larger than $\ell_{\mathrm{large}}$ is dominant, whereas the contribution from scales much smaller than $\ell_{\mathrm{large}}$ is negligible.
The existence of such a characteristic length scale is ensured by the decay of the pressure-dilatation co-spectrum at a large $k$, which is well established for ordinary compressible turbulence \cite{Aluie_2012,Wang-Yang-Shi}:
\begin{equation}
E^{(p)}(k)=O(k^{-\alpha}),\quad \alpha>1,
\label{eq:E^PD condition}
\end{equation}
where $E^{(p)}(k)$ is defined by
\begin{equation}
E^{(p)}(k):=-\dfrac{1}{\Delta k}\sum_{k-\Delta k/2<|{\bf k}|<k+\Delta k/2}\hat{p}({\bf k})\widehat{\nabla\cdot{\bf v}}(-{\bf k}).
\end{equation}
Here, $\Delta k:=2\pi/\mathcal{L}$.
Using the pressure-dilatation co-spectrum, the characteristic length scale $\ell_{\mathrm{large}}$ is explicitly defined, for instance, as
\begin{eqnarray}
\ell_{\mathrm{large}}:=\dfrac{\sum_kk^{-1}E^{(p)}(k)}{\sum_kE^{(p)}(k)}.
\label{ell_large}
\end{eqnarray}
From (\ref{eq:E^PD condition}) and (\ref{ell_large}), it follows that the mean large-scale pressure-dilatation $\bra\bar{p}_\ell\nabla\cdot\bar{{\bf v}}_\ell\ket$ converges to the finite constant $\bra p\nabla\cdot{\bf v}\ket$ and becomes independent of $\ell$ at scales sufficiently smaller than $\ell_{\mathrm{large}}$; this is expressed as
\begin{eqnarray}
\bra p\nabla\cdot{\bf v}\ket&=&-\lim_{K\rightarrow\infty}\sum_{0\le k<K}E^{(p)}(k)\notag\\
&\approx&-\sum_{0\le k<\ell^{-1}_{\mathrm{large}}}E^{(p)}(k)\notag\\
&\approx&\bra\bar{p}_\ell\nabla\cdot\bar{{\bf v}}_\ell\ket\quad\text{for}\quad\ell\ll\ell_{\mathrm{large}}.
\label{eq:saturation PD}
\end{eqnarray}

In the main text, $\ell_{\mathrm{small}}$ is introduced as the characteristic length scale, such that the contribution to the global vdW-stress--strain $\bra-{\bf \Sigma}:\nabla{\bf v}\ket$ from scales much larger than $\ell_{\mathrm{small}}$ is negligible whereas the contribution from scales much smaller than $\ell_{\mathrm{small}}$ is dominant.
The existence of such a characteristic length scale is validated using (\ref{estimation_Koretweg-stress-strain}).
The characteristic length scale $\ell_{\mathrm{small}}$ is explicitly defined, for instance, as
\begin{equation}
\ell_{\mathrm{small}}:=\dfrac{\sum_kk^{-1}E^{(\Sigma)}(k)}{\sum_kE^{(\Sigma)}(k)},
\label{ell_small}
\end{equation}
where $E^{(\Sigma)}(k)$ is the vdW-stress--strain co-spectrum defined by
\begin{equation}
E^{(\Sigma)}(k):=-\dfrac{1}{\Delta k}\sum_{k-\Delta k/2<|{\bf k}|<k+\Delta k/2}\hat{{\bf \Sigma}}({\bf k}):\widehat{\nabla{\bf v}}(-{\bf k}).
\end{equation}
From (\ref{estimation_Koretweg-stress-strain}) and (\ref{ell_small}), it follows that the mean large-scale vdW-stress--strain, $\bra-\bar{{\bf \Sigma}}_\ell:\nabla\bar{{\bf v}}_\ell\ket\le\bra|\bar{{\bf \Sigma}}_\ell:\nabla\bar{{\bf v}}_\ell|\ket=\|\bar{{\bf \Sigma}}_\ell:\nabla\bar{{\bf v}}_\ell\|_1$, is negligible at scales sufficiently larger than $\ell_{\mathrm{small}}$; this is expressed as
\begin{eqnarray}
\bra-\bar{{\bf \Sigma}}_\ell:\nabla\bar{{\bf v}}_\ell\ket\approx0\quad\text{for}\quad\ell\gg\ell_{\mathrm{small}}.
\end{eqnarray}

Combining these results, (\ref{eq:Q}) becomes
\begin{eqnarray}
\bra Q^{\mathrm{flux}}_\ell\ket&\approx&\bra p\nabla\cdot{\bf v}\ket+\bra{\bf v}\cdot{\bf f}\ket\notag\\
&=:&\epsilon_{\mathrm{eff}}\quad\text{for}\quad\ell_{\mathrm{small}}\ll\ell\ll\ell_{\mathrm{large}}.
\label{eq:Q_inertial_range}
\end{eqnarray} 
Because $\bra Q^{\mathrm{flux}}_\ell\ket$ can be expressed in terms of increments, as shown in Sec. \ref{Scale dependence of energy fluxes and vdW-stress--strain}, (\ref{eq:Q_inertial_range}) plays the same role as Kolmogorov's $4/5$-law.

\subsection{Existence of the van der Waals cascade\label{Existence of the van der Waals cascade}}
\subsubsection{Basis of the estimation $\bar{\Sigma}_\ell\delta\rho(\ell)\sim Z$}
Before we explain the existence of the van der Waals cascade using the results in Sec.\ \ref{Scale dependence of energy fluxes and vdW-stress--strain}, we explain the basis of the estimation $\bar{\Sigma}_\ell\delta\rho(\ell)\sim Z$, which is used in the main text.
% \subsubsection{Explanation of the relation $\bar{{\bf \Sigma}}_\ell\delta\rho(\ell)\sim Z$}
From assumption (\ref{assumption_rho}) and the estimation (\ref{eq:estimation_Sigma}), we obtain
\begin{eqnarray}
\|\bar{{\bf \Sigma}}_\ell\delta\rho(\ell)\|_p&\le&\|\bar{{\bf \Sigma}}_\ell\|_\infty\|\delta\rho(\ell)\|_p\notag\\
&=&O(1)
\end{eqnarray}
for all $p\in[1,\infty]$.
This evaluation is the basis of the estimation, $\bar{\Sigma}_\ell\delta\rho(\ell)\sim Z$.

\subsubsection{Explanation of (22)}
Here, using the results in Sec.\ \ref{Scale dependence of energy fluxes and vdW-stress--strain}, we explain (22), i.e.,
\begin{equation}
\bra Q^{\mathrm{flux}}_\ell\ket\approx
\begin{cases}
\bra\Pi_\ell\ket+\bra\Lambda^{(p)}_\ell\ket\approx\epsilon_{\mathrm{eff}}\quad\text{for}\quad\ell_c\ll\ell\ll\ell_{\mathrm{large}},\\
\bra\Lambda^{(\Sigma)}_\ell\ket\approx\epsilon_{\mathrm{eff}}\quad\text{for}\quad\ell_{\mathrm{small}}\ll\ell\ll\ell_c.
\end{cases}
\label{eq:Q_double-cascade}
\end{equation}
From (\ref{eq:Estimation of the second term}) and (\ref{estimation_Capillary_work}), it immediately follows that the upper bounds of the mean deformation work, $\bra\Pi_\ell\ket$, and mean capillary work, $\bra\Lambda^{(\Sigma)}_\ell\ket$, have different $\ell$ dependences.
In particular, in the case of $\sigma_3=1/3$,
\begin{equation}
\bra\Pi_\ell\ket\le\bra|\Pi_\ell|\ket=\|\Pi_\ell\|_1=O(1),
\end{equation}
\begin{equation}
\bra\Lambda^{(\Sigma)}_\ell\ket\le\bra|\Lambda^{(\Sigma)}_\ell|\ket=\|\Lambda^{(\Sigma)}_\ell\|_1=O\left(\left(\dfrac{\ell}{L}\right)^{-2/3}\right),
\end{equation}
whereas in the case of $\sigma_3=1$,
\begin{equation}
\bra\Pi_\ell\ket\le\bra|\Pi_\ell|\ket=\|\Pi_\ell\|_1=O\left(\left(\dfrac{\ell}{L}\right)^2\right),
\end{equation}
\begin{equation}
\bra\Lambda^{(\Sigma)}_\ell\ket\le\bra|\Lambda^{(\Sigma)}_\ell|\ket=\|\Lambda^{(\Sigma)}_\ell\|_1=O(1).
\end{equation}

\begin{figure}[t]
\includegraphics[width=8.6cm]{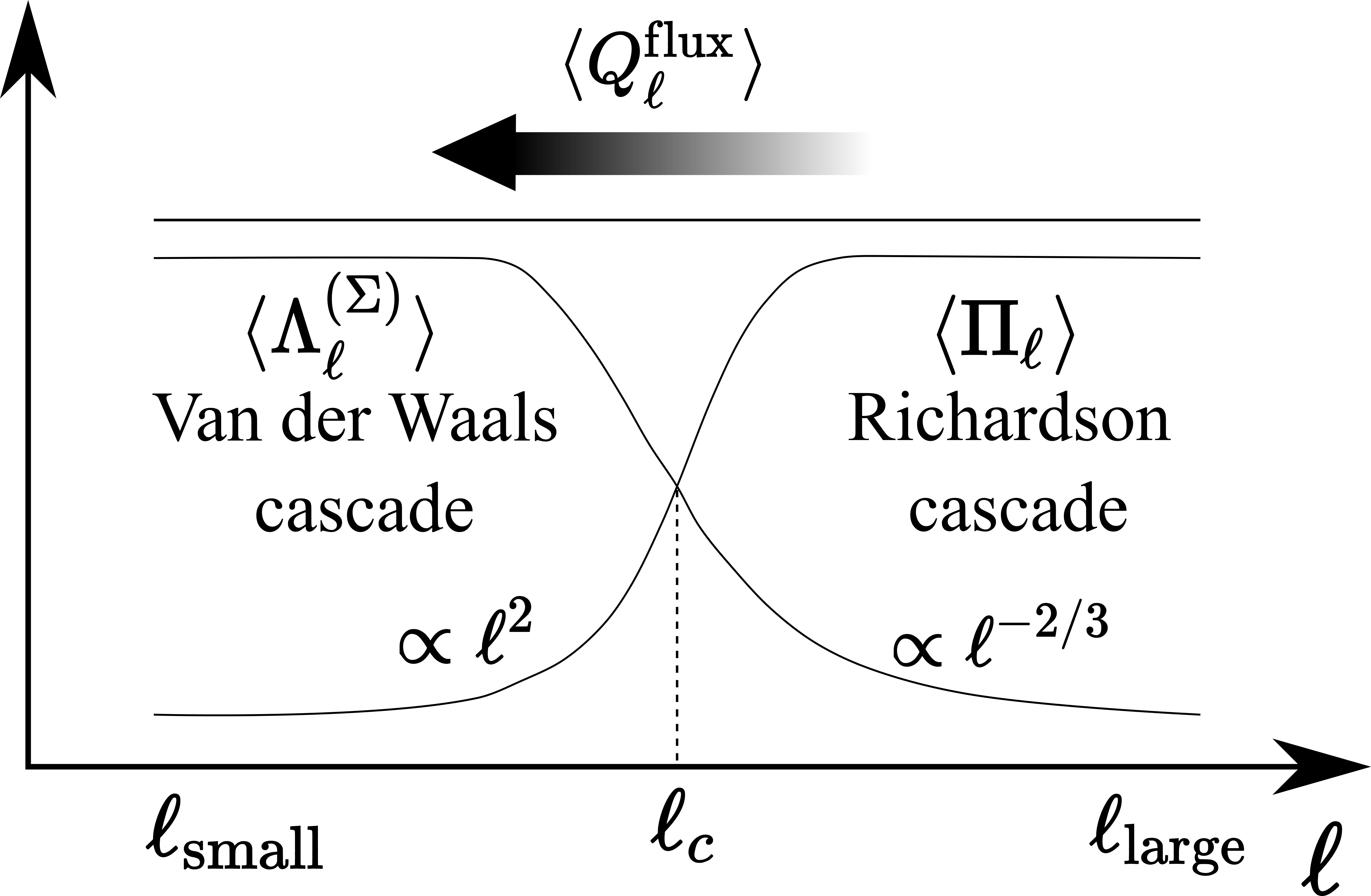}
\caption{Scale dependence of the scale-to-scale kinetic energy fluxes. The solid lines indicate the upper bounds of the energy fluxes, and the arrow indicates the direction of energy transfer.}
\label{fig:crossover}
\end{figure}

The ``Kolmogorov's $4/5$-law'' states that the sum of the mean deformation work and mean capillary work, $\bra\Pi_\ell\ket+\bra\Lambda^{(\Sigma)}_\ell\ket$, becomes scale-independent in the inertial range $\ell_{\mathrm{small}}\ll\ell\ll\ell_{\mathrm{large}}$.
From this law and the above observation, if we ignore the contribution of baropycnal work based on the evaluation (\ref{eq:Estimation of baropycnal work}), it follows that a characteristic length scale $\lambda$ exists such that the energy cascade due to the deformation work is dominant in $\lambda\ll\ell\ll\ell_{\mathrm{large}}$, whereas that due to capillary work is dominant in $\ell_{\mathrm{small}}\ll\ell\ll\lambda$ (see Fig.~\ref{fig:crossover}).
This is expressed as follows:
\begin{eqnarray}
\bra\Lambda^{(\Sigma)}_\ell\ket\ll\bra\Pi_\ell\ket&=&O(1)\quad\text{for}\quad\lambda\ll\ell\ll\ell_{\mathrm{large}},\\
\bra\Pi_\ell\ket\ll\bra\Lambda^{(\Sigma)}_\ell\ket&=&O(1)\quad\text{for}\quad\ell_{\mathrm{small}}\ll\ell\ll\lambda.
\end{eqnarray}
% The crossover length scale $\lambda$ may be determined, for instance, as follows:
From the definition of $\ell_c$, we expect that the crossover length scale $\lambda$ is of the order of $\ell_c$.
In fact, if we use an estimation that 
\begin{equation}
\Pi_\lambda=-\bar{\rho}_\lambda\nabla\tilde{{\bf v}}_\lambda:\tilde{\tau}_\lambda({\bf v},{\bf v})=\mathcal{O}\left(\rho_0\lambda^{-1}v^3_0\right),
\end{equation}
and
\begin{equation}
\Lambda^{(\Sigma)}_\lambda=(1/\bar{\rho}_\lambda)\nabla\cdot\bar{{\bf \Sigma}}_\lambda\cdot\bar{\tau}_\lambda(\rho,{\bf v})=\mathcal{O}\left(\lambda^{-3}T_0|c_0|\rho^2_0v_0\right),
\end{equation}
where the symbol $\mathcal{O}$ denotes ``same order of magnitude as,'' we obtain
\begin{eqnarray}
% \Pi_\lambda&\sim&\Lambda^{(\Sigma)}_\lambda\notag\\
% \Leftrightarrow\quad\lambda&\sim&\dfrac{\sqrt{T_0c_0\rho_0}}{v_0}.
\lambda&=&\mathcal{O}\left(\dfrac{\sqrt{T_0|c_0|\rho_0}}{v_0}\right)\notag\\
&=&\mathcal{O}\left(\ell_c\right).
\end{eqnarray}

Thus, a two type of cascade occurs in the van der Waals turbulence, one in $\ell_c\ll\ell\ll\ell_{\mathrm{large}}$ and the other in $\ell_{\mathrm{small}}\ll\ell\ll\ell_c$.
The former is the Richardson cascade, which is induced by the deformation work, as in the case of ordinary turbulence.
The latter is the van der Waals cascade, which is induced by capillary work, and its existence is specific to van der Waals turbulence.

\subsubsection{Velocity power spectrum}
Here, we explain the detailed derivation of the velocity power spectrum obtained in the main text.
In compressible turbulence, we can consider the spectra of the velocity ${\bf v}$ and the density-weighted velocity, such as $\sqrt{\rho}{\bf v}$ \cite{Kida_1990}.
In an ordinary compressible turbulence, high-resolution numerical simulations exhibit the Kolmogorov spectrum for both velocity \cite{Aluie_2012} and density-weighted velocity power spectra \cite{Wang-Yang-Shi} in the case where $\ell_{\mathrm{large}}$ is sufficiently larger than the Kolmogorov scale.
In this subsection, we consider the spectra of both the velocity ${\bf v}$ and the density-weighted velocity $\sqrt{\rho}{\bf v}$.

\paragraph*{Velocity power spectrum.}
First, we consider the $p$th-order (absolute) structure function for the velocity field,
\begin{equation}
S^v_p(\ell):=\bra|\delta{\bf v}(\ell)|^p\ket=\|\delta{\bf v}(\ell)\|^p_p
\label{S_p}
\end{equation} 
with assumed scaling exponent $\zeta_p$:
\begin{equation}
S^v_p(\ell)\sim C_pv^p_0\left(\dfrac{\ell}{L}\right)^{\zeta_p}\quad\text{as}\quad\ell/L\rightarrow0,
\label{S_p scaling}
\end{equation}
where $C_p$ is a dimensionless constant.
Using the H\"older inequality, it can be shown that $\zeta_p$ is a concave function of $p\in[0,\infty)$ \cite{Frisch,Eyink_lecture}.
From this property, it immediately follows that $\sigma_p=\zeta_p/p$ is a non-increasing function of $p$ \cite{Eyink_lecture}.
Note that the second-order structure function $S^v_2(\ell)\propto\ell^{\zeta_2}$ is related to the velocity spectrum $E^v(k)\propto k^{-\zeta_2-1}$, assuming isotropy.

Because $\sigma_3=1/3$ in $\ell_c\ll\ell\ll \ell_{\mathrm{large}}$ and $\sigma_p$ is a non-increasing function of $p$, it follows that $\sigma_2\ge1/3$ in this scale range.
Hence, we can write $\zeta_2=2\sigma_2\equiv2/3+\mu/9$, where $\mu$ is a positive constant.
This additional constant $\mu$ corresponds to the so-called \textit{intermittency exponent} \cite{Frisch}.
Therefore, the velocity power spectrum exhibits the following asymptotic behavior:
\begin{equation}
E^v(k)\sim C_{\mathrm{large}} k^{-5/3-\mu/9}\quad\text{for}\quad \ell^{-1}_{\mathrm{large}}\ll k\ll\ell^{-1}_c,
\label{eq:energy spectrum bound}
\end{equation}
where $C_{\mathrm{large}}$ is a positive constant.

In $\ell_{\mathrm{small}}\ll\ell\ll\ell_c$, where the van der Waals cascade becomes dominant, we have seen that $\sigma_3=1$.
Because $\sigma_p$ is a non-increasing function of $p$, it follows that $\sigma_2=1$.
This result implies that the velocity power spectrum exhibits the following asymptotic behavior:
\begin{equation}
E^v(k)\sim C_{\mathrm{small}}k^{-3}\quad\text{for}\quad \ell^{-1}_c\ll k\ll\ell^{-1}_{\mathrm{small}},
\label{eq:energy spectrum bound_2}
\end{equation}
where $C_{\mathrm{small}}$ is a positive constant.

\begin{figure}[t]
\includegraphics[width=8.6cm]{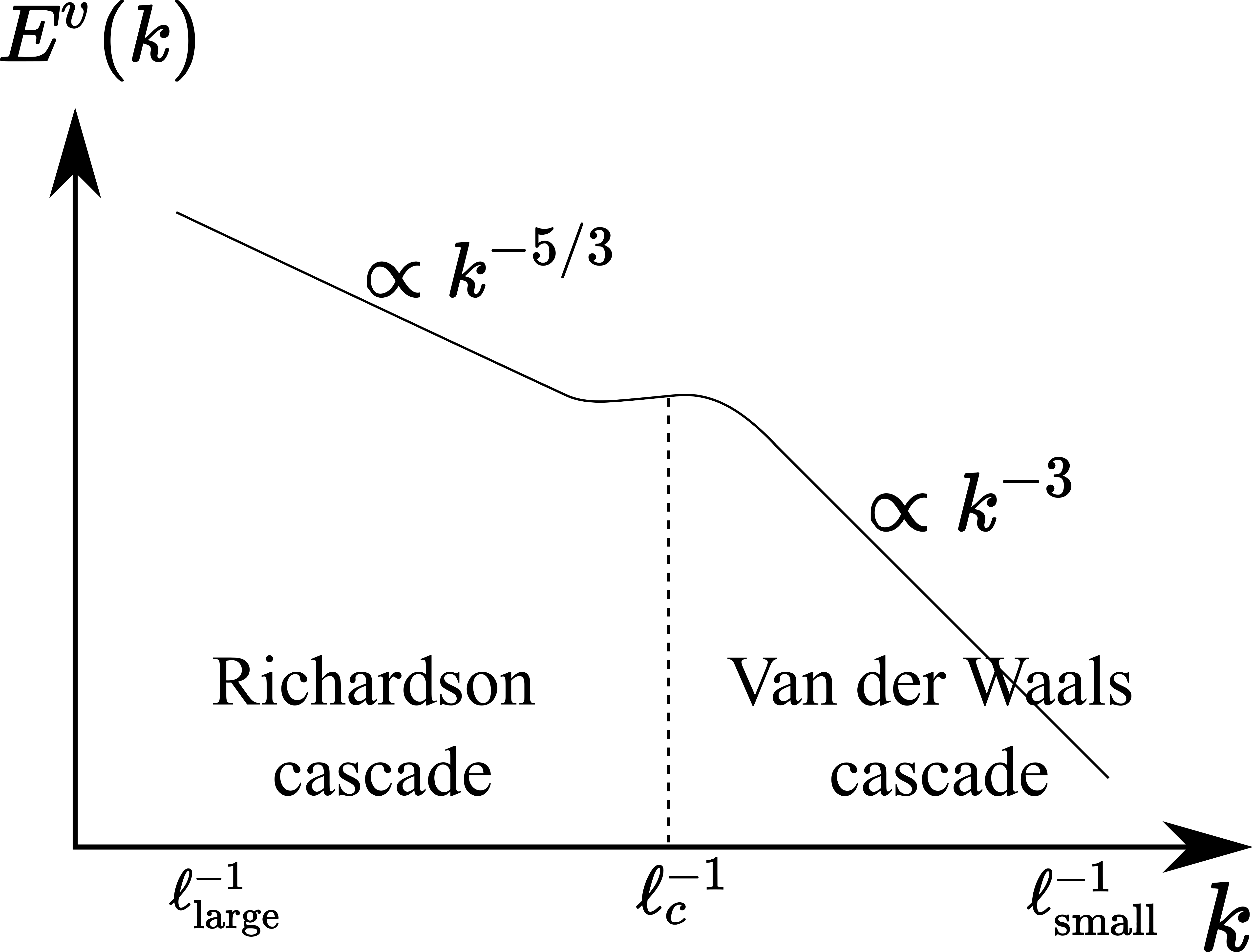}
\caption{Velocity power spectrum $E^v(k)$ in the inertial range $\ell^{-1}_{\mathrm{large}}\ll k\ll\ell^{-1}_{\mathrm{small}}$.}
\label{fig:energy_spectrum}
\end{figure}

This result is summarized in Fig. \ref{fig:energy_spectrum}.

\paragraph*{Density-weighted velocity power spectrum.}
Next, we consider the spectrum of the density-weighted velocity $\sqrt{\rho}{\bf v}$.
We consider the $p$th-order (absolute) structure function for the density-weighted velocity,
\begin{equation}
S^{\sqrt{\rho}v}_p(\ell):=\bra|\delta(\sqrt{\rho}{\bf v})(\ell)|^p\ket=\|\delta(\sqrt{\rho}{\bf v})(\ell)\|^p_p,
\end{equation} 
with an assumed scaling exponent, $\tilde{\zeta}_p$.
\begin{equation}
S^{\sqrt{\rho}v}_p(\ell)\sim \tilde{C}_p\rho^{p/2}v^p_0\left(\dfrac{\ell}{L}\right)^{\tilde{\zeta}_p}\quad\text{as}\quad\ell/L\rightarrow0,
\end{equation}
where $\tilde{C}_p$ is a dimensionless constant.
Using the H\"older inequality, we can also see that $\tilde{\zeta}_p$ is a concave function of $p\in[0,\infty)$ \cite{Frisch,Eyink_lecture}.
Note that the second-order structure function, $S^{\sqrt{\rho}v}_2(\ell)\propto\ell^{\tilde{\zeta}_2}$, is also related to the density-weighted velocity spectrum, $E(k)\propto k^{-\tilde{\zeta}_2-1}$, assuming isotropy.

In this case, we cannot determine the exact value of $\tilde{\zeta}_2$ because, from the mean value theorem, Minkowski inequality, and assumptions (\ref{assumption_v}) and (\ref{assumption_rho}),
\begin{eqnarray}
\|\delta(\sqrt{\rho}{\bf v})({\bf r};\cdot)\|_p&\le&B_1\|\delta\rho({\bf r};\cdot)\|_p+B_2\|\delta{\bf v}({\bf r};\cdot)\|_p\notag\\
&\sim&B_3v_0\left(\dfrac{|{\bf r}|}{L}\right)^{\sigma_p}\quad\text{as}\quad|{\bf r}|/L\rightarrow0,\notag
\end{eqnarray}
where $B_1$, $B_2$, and $B_3$ are constants.
Hence, 
\begin{equation}
S^{\sqrt{\rho}v}_p(\ell)=\|\delta(\sqrt{\rho}{\bf v})(\ell)\|^p_p=O\left(\left(\dfrac{\ell}{L}\right)^{p\sigma_p}\right),
\end{equation}
and we cannot conclude that $\tilde{\zeta}_p=p\sigma_p$ in general.
If we assume that $\zeta_2\approx\tilde{\zeta}_2$, as in an ordinary compressible turbulence, the asymptotic behavior of the kinetic energy spectrum $E(k)$ can be obtained as
\begin{equation}
E(k)\sim 
\begin{cases}
\tilde{C}_{\mathrm{large}} k^{-5/3-\mu/9}\quad\text{for}\quad \ell^{-1}_{\mathrm{large}}\ll k\ll\ell^{-1}_c,\\
\tilde{C}_{\mathrm{small}}k^{-3}\quad\text{for}\quad \ell^{-1}_c\ll k\ll\ell^{-1}_{\mathrm{small}},
\end{cases}
\end{equation}
where $\tilde{C}_{\mathrm{large}}$ and $\tilde{C}_{\mathrm{small}}$ are positive constants.

\subsection{Additional assumption on the temperature and density gradient field}
In this section, we show that, if we further assume the regularity of the temperature and density gradient field, the evaluation of the capillary work (\ref{estimation_Capillary_work}) is not optimal, and the spectrum $\propto k^{-3}$ becomes shallower.
To this end, we additionally assume the following:
\begin{eqnarray}
\|\delta(\nabla\rho)({\bf r};\cdot)\|_p&=&O\left(\left(\dfrac{|{\bf r}|}{L}\right)^{\sigma^{\nabla\rho}_p}\right),\label{additional_assumption_nabla_rho}\\
\|\delta T({\bf r};\cdot)\|_p&=&O\left(\left(\dfrac{|{\bf r}|}{L}\right)^{\sigma^T_p}\right),\label{additional_assumption_T}
\end{eqnarray}
where $\sigma^{\nabla\rho}_p,\sigma^T_p\in[0,1)$.
Then, the evaluation of $\bar{{\bf \Sigma}}_\ell$ is modified as follows.
\begin{eqnarray}
\overline{(Tc\rho\Delta\rho)}_\ell&=&-\dfrac{1}{\ell}\int_\Omega d^3{\bf r}(\nabla G)_\ell({\bf r})\cdot(Tc\rho\nabla\rho)({\bf x}+{\bf r})-\int_\Omega d^3{\bf r}G_\ell({\bf r})(\nabla(Tc\rho)\cdot\nabla\rho)({\bf x}+{\bf r})\notag\\
% &=&-\dfrac{1}{\ell}\int_\Omega d^3{\bf r}(\nabla G)_\ell({\bf r})\cdot\delta(Tc\rho\nabla\rho)({\bf r};{\bf x})-\int_\Omega d^3{\bf r}G_\ell({\bf r})\delta(\nabla(Tc\rho)\cdot\nabla\rho)({\bf r};{\bf x})
&=&-\dfrac{1}{\ell}\int_\Omega d^3{\bf r}(\nabla G)_\ell({\bf r})\cdot\delta(Tc\rho\nabla\rho)({\bf r};{\bf x})-\int_\Omega d^3{\bf r}G_\ell({\bf r})(\nabla(Tc\rho)\cdot\nabla\rho)({\bf x}+{\bf r}),
\end{eqnarray}
and 
\begin{eqnarray}
\|\delta(Tc\rho\nabla\rho)({\bf r};\cdot)\|_p&=&\left\|\left.\dfrac{\partial (Tc\rho\nabla\rho)}{\partial T}\right|_{(T,\rho,\nabla\rho)=(T_*,\rho_*,\nabla\rho_*)}\delta T({\bf r};\cdot)+\left.\dfrac{\partial (Tc\rho\nabla\rho)}{\partial \rho}\right|_{(T,\rho,\nabla\rho)=(T_*,\rho_*,\nabla\rho_*)}\delta \rho({\bf r};\cdot)\right.\notag\\
&&+\left.\left.\dfrac{\partial (Tc\rho\nabla\rho)}{\partial \nabla\rho}\right|_{(T,\rho,\nabla\rho)=(T_*,\rho_*,\nabla\rho_*)}\delta(\nabla\rho)({\bf r};\cdot)\right\|_p\notag\\
&=&O\left(\left(\dfrac{|{\bf r}|}{L}\right)^{\min\{\sigma^{\nabla\rho}_p,\sigma^T_p\}}\right),
\end{eqnarray}
% \begin{eqnarray}
% \|\delta(\nabla(Tc\rho)\cdot\nabla\rho)({\bf r};\cdot)\|_p
% \end{eqnarray}
where $(T_*,\rho_*,\nabla\rho_*)$ is on the line segment joining $(T({\bf x}),\rho({\bf x}),\nabla\rho({\bf x}))$ and $(T({\bf x}+{\bf r}),\rho({\bf x}+{\bf r}),\nabla\rho({\bf x}+{\bf r}))$.
Thus, it follows that
\begin{eqnarray}
\|\overline{(Tc\rho\Delta\rho)}_\ell\|_p&\le&\dfrac{1}{\ell}\int_\Omega d^3{\bf r}|(\nabla G)_\ell({\bf r})|\|\delta(Tc\rho\nabla\rho)({\bf r};\cdot)\|_p+\int_\Omega d^3{\bf r}G_\ell({\bf r})\|(\nabla(Tc\rho)\cdot\nabla\rho)\|_p\notag\\
&\le&\dfrac{\|\delta(Tc\rho\nabla\rho)(\ell)\|_p}{\ell}\int_\Omega d^3{\bf r}|(\nabla G)_\ell({\bf r})|+\|(\nabla(Tc\rho)\cdot\nabla\rho)\|_p\notag\\
&=&O\left(\left(\dfrac{\ell}{L}\right)^{\min\{\sigma^{\nabla\rho}_p,\sigma^T_p\}-1}\right).
\end{eqnarray}

Therefore, $\|\bar{{\bf \Sigma}}_\ell\|_p=O(\ell^{\min\{\sigma^{\nabla\rho}_p,\sigma^T_p\}-1})$, and the estimations of capillary work and the vdW-stress--strain are modified as follows:
\begin{eqnarray}
\|\Lambda^{(\Sigma)}_\ell\|_{p/3}&=&O\left(\left(\dfrac{\ell}{L}\right)^{\sigma_p+\min\{\sigma^{\nabla\rho}_p,\sigma^T_p\}-1}\right),\\
\|\bar{{\bf \Sigma}}_\ell:\nabla\bar{{\bf v}}_\ell\|_p&=&O\left(\left(\dfrac{\ell}{L}\right)^{\sigma_p+\min\{\sigma^{\nabla\rho}_p,\sigma^T_p\}-2}\right).
\end{eqnarray}
Hence, the asymptotic behavior of the velocity power spectrum can be evaluated as
\begin{equation}
E^v(k)\propto k^{-3+2\min\{\sigma^{\nabla\rho}_p,\sigma^T_p\}}\quad\text{for}\quad\ell^{-1}_c\ll k\ll\ell^{-1}_{\mathrm{small}}.
\end{equation}
Note that the spectral index value $-3$ corresponds to the case of either $\sigma^{\nabla\rho}_p$ or $\sigma^T_p$ equaling zero.

\bibliography{supplemental_material}